\documentclass[12pt]{article}
\usepackage{graphics,cite,amssymb,epsfig,float}
\usepackage[usenames,dvips]{color}
\usepackage{citesort}
\usepackage{amsmath, mathbbol}

\makeatletter

\@addtoreset{equation}{Section}
\makeatother

\def\lsim{\;\raise0.3ex\hbox{$<$\kern-0.75em\raise-1.1ex\hbox{$\sim$}}\;}
\def\gsim{\;\raise0.3ex\hbox{$>$\kern-0.75em\raise-1.1ex\hbox{$\sim$}}\;}

\def\ben{\begin{enumerate}}  \def\een{\end{enumerate}}
\def\bit{\begin{itemize}}    \def\eit{\end{itemize}}
\def\beq{\begin{equation}}   \def\eeq{\end{equation}}
\def\ba{\begin{array}}       \def\ea{\end{array}}
\def\bea{\begin{eqnarray}}   \def\eea{\end{eqnarray}}
\def\nn{\nonumber}

\begin{document}

\begin{titlepage}
\renewcommand{\thefootnote}{\fnsymbol{footnote}}
\setcounter{footnote}{0}

\vspace*{-2cm}
\begin{flushright}
LPT Orsay 08-82 \\
\today
\end{flushright}

\begin{center}
\vspace{2cm}
{\Large\bf Phenomenology of the constrained NMSSM} \\
\vspace{1cm}
{\bf A. Djouadi, U. Ellwanger and A.M. Teixeira\footnote{
Address after 01/11/2008:  CFTP, Instituto Superior T\'ecnico, Av. 
Rovisco Pais 1, 1049-001 Lisboa, Portugal.} } \\ 
 \vspace*{.5cm} 
Laboratoire de Physique Th\'eorique, CNRS -- UMR 8627, \\
Universit\'e de Paris XI, F-91405 Orsay Cedex, France
\vspace{2cm}
\end{center}

\begin{abstract}

We discuss several phenomenological aspects of the fully constrained
version of the next-to-minimal supersymmetric extension of the standard
model (cNMSSM).  Assuming universal boundary conditions at a high energy
scale for the soft supersymmetry-breaking gaugino, sfermion and Higgs
masses as well as for the trilinear interactions, we find that
the model can satisfy all present constraints from colliders and
cosmological data on dark matter, $B$- and muon-physics. The
phenomenologically viable region of the parameter space of the cNMSSM can
be described by essentially one single parameter as the universal gaugino
mass parameter $M_{1/2}$, and  corresponds to small values for the
universal scalar mass $m_0$. The lightest supersymmetric particle is
always a  singlino-like neutralino that is almost degenerate with 
the lightest tau slepton.   We study the particle spectrum of the
model and its signatures at the LHC, such as possibly long-lived tau
sleptons at the end of decay chains, that would distinguish the cNMSSM
from the constrained MSSM. 

\end{abstract}

\end{titlepage}

\renewcommand{\thefootnote}{\arabic{footnote}}
\setcounter{footnote}{0}

\vspace*{5mm}
\section{Introduction} 

The next-to-minimal supersymmetric standard model (NMSSM)
\cite{genNMSSM1,genNMSSM2,genNMSSM3}, in which the spectrum of the
minimal extension (MSSM) is extended by one singlet superfield, was among
the first supersymmetric (SUSY) models based on supergravity-induced
SUSY-breaking terms. It has gained a renewed interest in the last decade,
since it solves in a natural and elegant way the so-called $\mu$
problem~\cite{MuProblem} of the MSSM: in the NMSSM, this parameter is
linked to the vacuum expectation value (vev) of the singlet Higgs field,
and  a value close to the SUSY-breaking scale is dynamically generated. 
In fact, the NMSSM is the simplest supersymmetric generalization of the
standard model (SM) in which the SUSY breaking scale is the  {\em only}
scale in the Lagrangian, since it allows for a scale invariant
superpotential.

In contrast to the non or partially constrained versions of the NMSSM
that have been intensively studied in recent years~\cite{benchmark}  and
which involve many free parameters, the constrained model (cNMSSM) has
soft SUSY-breaking para\-meters that are universal at a high scale such
as the grand unification (GUT) scale: common gaugino ($M_{1/2}$) and
scalar ($m_0$) masses  as well as trilinear couplings ($A_0$),  as
motivated by schemes for SUSY-breaking that are mediated by flavour blind
gravitational interactions. If one assumes that the soft SUSY-breaking
para\-meters involving the additional singlet sector are  universal as
well, the model has the same number of unknown parameters as the
constrained MSSM (cMSSM)~\cite{msugra}.  

General features of the  constrained NMSSM parameter space as well as
aspects of its  phenomenology have been discussed earlier in
Refs.~\cite{genNMSSM2,genNMSSM3}. These studies already revealed that the
allowed range for the parameters $M_{1/2},\; m_0$ and $A_0$ is different
from that of the cMSSM. Small values for $m_0$ are disfavored in
the cMSSM, as they lead to charged sleptons that are lighter than the
neutralino $\chi_1^0$, the preferred lightest SUSY particle
(LSP). In the cNMSSM, small $m_0$ is needed to generate a non-vanishing
vev of the singlet Higgs field \cite{genNMSSM2}. The slepton LSP problem
can be evaded owing to the presence of the additional singlino-like
neutralino which, in large regions of the cNMSSM parameter space, is the
true LSP~{\cite{hugbelpuk}}.

Since the early studies of the cNMSSM,  bounds on the Higgs and SUSY
particle spectrum from high-energy collider data and low-energy
measurements have become more severe, and the dark matter relic density
has been determined quite accurately~\cite{wmap}. In addition, tools for
a more precise determination of the mass spectrum and couplings
\cite{nmssmtools}, and for the computation of the dark matter relic
density \cite{belsemenov}, have become available.  In fact, the tools
which allow to calculate the spectrum in the fully constrained NMSSM have
only been developed recently and a brief account of the resulting
phenomenology has been reported in Ref.~\cite{Djouadi:2008yj}. In this
paper, we investigate in more detail the parameter space of the cNMSSM
in light  of the recent constraints, using the updated tools. 

A priori, it is not obvious if the latest constraints on the Higgs sector
from LEP~\cite{Schael:2006cr} and on the dark matter relic
density~\cite{wmap} can be simultaneously satisfied in the  fully
constrained cNMSSM with its additional CP-even and CP-odd Higgs states
and a singlino-like LSP.  By scanning the cNMSSM parameter space with
with the program NMSSMTools~\cite{nmssmtools}, which calculates the 
Higgs and SUSY particle spectra in the NMSSM, we found regions where
this  is indeed possible~\cite{Djouadi:2008yj}. These phenomenologically 
viable regions correspond to a regime of very small or vanishing
universal scalar mass $m_0$, and the requirement of a correct
relic density determines the universal  trilinear $A_0$ as a function of
the universal gaugino mass  $M_{1/2}$. For low values of $M_{1/2}$, for
which $A_0$ is also quite small, the soft SUSY-breaking terms are
necessarily close to their ``no-scale'' values, $A_0 = m_0 = 0$
\cite{noscale}. ($m_0 = 0$, but nonvanishing values for $M_{1/2}$
and $A_0$ at a high scale, can also originate from a strongly
interacting conformal hidden sector~\cite{scalarseq}). For small enough
values of the
Yukawa coupling between the two doublet and the singlet Higgs
fields, $\lambda \lsim 10^{-2}$,  constraints from  LEP on the Higgs
sector as well as constraints from $B$-physics are also satisfied.
Notably, the model is very predictive in the sense that the complete
sparticle spectrum depends essentially on just one parameter, which can
be taken as $M_{1/2}$.

Moreover, in these viable regions of the parameter space,  the cNMSSM
can also explain  the deviation of the observed anomalous magnetic moment
of the muon $(g-2)_\mu$ with respect to the SM expectation~\cite{g-2}
{\em as well as} the $1.7\sigma$ and $2.3\,\sigma$ excesses of events
observed at  LEP~\cite{LEP98excess,Schael:2006cr}
corresponding to Higgs masses around 115 GeV and 98~GeV, respectively.

Some of these results have already been published 
in Ref.~\cite{Djouadi:2008yj}; here we discuss in more detail the full
allowed parameter space, the complete sparticle and Higgs spectrum, 
as well as the phenomenological implications for future experiments,
notably for the LHC. 

An unconventional but general property of the sparticle spectrum of the
cNMSSM is a singlino-like LSP with very small couplings to non-singlet
particles, and a stau next-to-LSP (NLSP) with a mass close to
that of the LSP.
The stau will appear in practically all sparticle decay
cascades. The small value of $m_0$ as compared to $M_{1/2}$ implies that
all squarks are  lighter than the gluino, a feature which will be
relevant for sparticle searches at the LHC. The SM-like CP-even Higgs
boson has a mass in the 115--120~GeV  range.

In some regions of the parameter space of the cNMSSM, for instance for a very 
small Yukawa coupling $\lambda$, the stau lifetime can
be very large, with  visibly displaced vertices originating from its
decay.  For larger values of $\lambda$, the additional mostly
singlet-like CP-even Higgs boson can have sizable couplings to gauge
bosons and a mass around 98~GeV, which would in turn explain the
$2.3\,\sigma$ excess of events observed at
LEP~\cite{LEP98excess,Schael:2006cr};  the other $1.7\,\sigma$
excess of  Higgs-like events would be due to the nearly SM-like
next-to-lightest CP-even Higgs with a mass close to $115$
GeV.  

Hence, not only the sparticle spectrum of the cNMSSM should already allow to
discriminate it from the cMSSM at the LHC, but additional unconventional
phenomena such as displaced vertices or a more complicated Higgs sector
can also occur.

The layout of the paper is as follows: in Section~2 we describe the model,
its free parameters and discuss the phenomenologically viable 
cNMSSM parameter space with the help of analytic approximations. 
In Section~3, we present the results for the
Higgs  and sparticle spectra. In Section~4 we discuss additional
phenomenological aspects of the model  that are relevant to the LHC:
sparticle production cross sections 
and decay cascades, the  possibility of displaced
vertices from stau decays and features of the Higgs sector. Finally we
comment on  tests at $e^+e^-$ colliders and  the direct and indirect 
detection of dark matter, which can rule out the present model.  A
summary and concluding remarks are presented in Section~5.

\section{The constrained NMSSM}

\subsection{The parameters of the cNMSSM}\label{param}
 
We consider the NMSSM with a scale invariant superpotential given by
\begin{equation} 
{\cal W}\!=\!
h_t \,\widehat{Q}\,\widehat{H}_u\,\widehat{t}_R^c - h_b \,\widehat{Q}
\,\widehat{H}_d\,\widehat{b}_R^c  - h_\tau \,\widehat{L}\, \widehat{H}_d
\,\widehat{\tau}_R^c +
\lambda \,\widehat{S}\, \widehat{H}_u\, \widehat{H}_d +
\frac{\kappa}{3}\, \widehat{S}^3
\;,  
\label{supot} 
\end{equation}
where hatted letters denote superfields. $H_u$, $H_d$ and $S$
represent the complex scalar Higgs fields, with $h_u$, $h_d$ and $s$ their
vacuum expectation values. Tilded letters will denote the scalar
components of quark and lepton superfields. For simplicity, only third
generation (s)fermions have been included and $\widehat Q, \widehat L$
stand for the $(t,b)$ and $(\tau,\nu_\tau)$
SU(2) doublet superfields. 
The three first terms in eq.~(\ref{supot}) are the usual generalization of
the Yukawa interactions, while the last two terms involving the singlet
superfield $\widehat{S}$ substitute the $\mu \widehat H_u  \widehat H_d$
term in the MSSM superpotential: a non-vanishing value $s$ at the
minimum of the Higgs potential generates an effective $\mu$ term
\beq
\mu_{\mathrm{eff}} \equiv \lambda\, s\; .
\eeq

The singlet superfield $\widehat{S}$ contains a neutral CP-even and a
neutral CP-odd scalar, as well as a neutralino. All these states 
mix with the corresponding components of the $\widehat{H}_u$ and
$\widehat{H}_d$ superfields, increasing the rank of the CP-even, CP-odd and
neutralino mass matrices by one as compared to the MSSM. Conventions for
signs and mixing matrices are chosen as in the SLHA2~\cite{slha2}
convention, and we take $\lambda>0$.

The soft SUSY--breaking terms consist of mass terms for the
gaugino, Higgs and sfermion  fields
(for the latter, we will use the notation of the third generation;
a sum over the three generations is implicitly assumed)
 \bea
-{\cal L}_\mathrm{\frac12}\!&\!=\!&\! \frac{1}{2} \bigg[ 
 M_1 \tilde{B}  \tilde{B}
\!+\!M_2 \sum_{a=1}^3 \tilde{W}^a \tilde{W}_a 
\!+\!M_3 \sum_{a=1}^8 \tilde{G}^a \tilde{G}_a   
\bigg]+ \mathrm{h.c.}\; , \\ 
 -{\cal L}_\mathrm{0} \!&\!=\!&\! 
m_{H_u}^2 | H_u |^2 + m_{H_d}^2 | H_d |^2 + 
m_{S}^2 | S |^2 +m_Q^2|\tilde Q^2| + m_t^2|\tilde t_R^2| \nn \\ &
&+\,m_b^2|\tilde b_R^2| +m_L^2|\tilde L^2|
+m_\mathrm{\tau}^2|\tilde\tau_R^2|\; ,
\eea
as well as trilinear interactions between the sfermion and the Higgs
fields, including the singlet field 
\bea
-{\cal L}_\mathrm{tril} \!&\!=\!&\! 
 \Bigl( h_t A_t\, \tilde Q\, H_u\, \tilde t_R^c
- h_b  A_b\, \tilde Q\, H_d \,\tilde b_R^c - h_\tau A_\tau\, \tilde L \,H_d
\,\tilde \tau_R^c \nn \\
\!& &\!+\,  \lambda A_\lambda\, H_u \,H_d \,S +  \frac{1}{3} \kappa 
 A_\kappa\,  S^3 \Bigl)+ \mathrm{h.c.}\;.
\eea

All parameters in the above Lagrangian depend on the energy scale via the
corresponding RG equations,  so that the dominant radiative corrections
involving large logarithms are  accounted for. In the fully constrained
cNMSSM, one imposes unification  of the soft SUSY--breaking gaugino
masses, sfermion and Higgs masses as well as   trilinear couplings at the
grand unification scale $M_{\rm GUT}$:
\bea
& & M_1  = M_2 = M_3 \equiv  M_{1/2}\, , \nn \\
& & m_{H_u}  =  m_{H_d} =
m_{S} = m_Q = m_t = m_b = m_L
= m_{\tau} \equiv m_0\, , \nn \\
& & A_t  =  A_b = A_\tau = A_\lambda = A_\kappa \equiv  A_0\, .
\eea

Then, apart from gauge and quark/lepton Yukawa couplings,  the
Lagrangian of the cNMSSM depends on the five input parameters
\beq\label{pars}
M_{1/2}\ ,   \ m_0 \ , \ A_0 \ , \ \lambda \ {\rm and} \  \kappa\;.
\eeq
Requiring the correct value of $M_Z$
reduces the dimension of the parameter space from five to four.

In principle, one could start with four independent parameters (such as
$m_0/M_{1/2}$, $A_0/M_{1/2}$, $\lambda$ and $\kappa$), and integrate the
RG equations for all soft terms from $M_{\rm GUT}$ down to the SUSY scale
$M_{\rm SUSY}$, defined by the order of magnitude of the soft SUSY-breaking
terms. Subsequently, one could minimize the effective potential with
respect to $h_u$, $h_d$ and $s$ and determine the overall scale of the
soft terms in eq.~(\ref{pars}) from the correct value of $M_{Z}$, 
as done in Ref.~\cite{genNMSSM2}. However, since $\tan\beta = h_u/h_d$
is then obtained as output (while the top quark Yukawa coupling $h_{t}$
would be an input), it becomes very difficult to obtain the correct value
for the top quark mass $m_t$  or, given $m_t$, $h_{t}$ can only be
obtained once $\tan\beta$ is known. Since $h_{t}$ is very important for
the radiative corrections and the RG evolution, it is much more
convenient to allow for $\tan\beta$ as an input parameter, which permits
to determine $h_{t}$ at the weak scale  from the beginning in terms
of~$m_{t}$.

All in all, the following procedure is feasible in practice: apart from
$M_Z$, the five parameters
\beq
M_{1/2}\ , \ m_0 \ , \ A_0 \ , \ \lambda \ {\rm and} \  \tan\beta\;
\eeq
are allowed as inputs. The parameters $\kappa$, the soft singlet mass
$m_S^2$ as well as the vev $|s|$ (or $|\mu_{\rm eff}| \equiv \lambda |s|$) are
determined at $M_{\rm SUSY}$ through the three minimization equations of
the scalar potential with respect to $h_u$, $h_d$ and $s$. (With the
convention $\lambda > 0$, $\kappa$ typically turns out to be positive as
well, and of $\mathcal{O}(\lambda/10)$; the sign of $s$ or $\mu_{\rm
eff}$ can still be chosen at will.) This is the procedure employed by
the routine NMSPEC within NMSSMTools \cite{nmssmtools}, which calculates
the spectra of the Higgs and SUSY particles in the NMSSM in terms of the
soft SUSY breaking terms at $M_{\rm GUT}$ (except for the parameter
$m_S^2$), $\tan\beta$ at the weak scale (defined by $M_Z$) and $\lambda$
at the SUSY scale $M_{\rm SUSY}$.

Clearly, the soft singlet mass squared $m_S^2$ at $M_{\rm GUT}$ will not
coincide with $m_0^2$ in general (for a recent analysis allowing for a
non-universal singlet mass term, see Ref.~\cite{Balazs:2008ph}). However,
one can confine oneself to regions in parameter space where the
difference between $m_S^2$ and $m_0^2$ is negligibly small. This
condition leaves us with an effective 4-dimensional parameter space,
consistent with the considerations above. In practice, we determine
$\tan\beta$ by the requirement that $m_S^2$ at $M_{\rm GUT}$ should be
close to $m_0^2$: we impose $|m_{S}^2(M_{\rm GUT}) - m_0^2| < (5~{\rm
GeV})^2$, which typically requires to tune the fourth decimal of
$\tan\beta$. This should not be interpreted as a fine-tuning, since
$m_S^2$ should be considered as an input parameter, whereas $\tan\beta$
is determined by the minimization of the effective potential. 

For the most relevant SM parameters, the strong coupling and the
bottom/top quark masses, we chose $\alpha_s(M_{Z}) = 0.1172$, $m_b^{
\overline{\rm MS}}(m_b) = 4.214$ GeV and $m_{{t}}^{\mathrm{pole}} =
171.4$ GeV \cite{pdg}.

\subsection{Constraints 
from the scalar potential}

Let us begin by recalling some conditions on the parameters $M_{1/2}$,
$m_0$ and $A_0$ of the \nobreak{cNMSSM}, which follow from a
phenomenologically acceptable minimum of the Higgs
potential~\cite{genNMSSM2}. First, the vev $s$ of the singlet has to be
non-vanishing. The dominant $s$-dependent terms in the Higgs potential
are given by
\beq\label{singpot}
V(s) \sim \kappa^2\, s^4 + \frac{2}{3}\kappa\, A_\kappa\, s^3 + m_S^2\,
s^2 + \dots\; ,
\eeq
and one easily finds that the condition for a non-vanishing
value for $s$ at the absolute minimum is equivalent to the inequality
\begin{equation}
m_S^2 \,\lesssim\, \frac{1}{9} A_\kappa^2\; .
\end{equation}

For small $\lambda$ and $\kappa$ (as will be the case below), the
parameters $A_\kappa$ and $m_S$ are hardly renormalized between the GUT
and the electroweak scales, and the above condition translates into the
first constraint on the parameter space (assuming $m_0^2 \geq 0$)
\begin{equation}\label{upperm0}
m_0 \,\lesssim\, \frac{1}{3} |A_0|\,.
\end{equation}

Next, we consider the CP-odd Higgs boson mass matrix. The dominant term in its
singlet-like diagonal element is given by 
\cite{genNMSSM1,genNMSSM2,genNMSSM3,nmssmtools,slha2}
\begin{equation}
-3 \kappa A_\kappa s, 
\end{equation}
and must be
positive. For positive $s$ and $\kappa$ this implies negative trilinear
couplings
\begin{equation}
A_\kappa\, \sim\, A_0 < 0\,.
\end{equation}

We must also consider the constraints on the parameter space arising from
vacuum stability. Dangerous instabilities  of the scalar potential along
charge and colour breaking (CCB) directions in field space can
occur~\cite{genNMSSM2,ccbnmssm}, notably for large values of $|A_0|$. The
most dangerous CCB direction is along the D-flat direction
\begin{equation}
|E_1|=|L_1|=|H_d| 
\end{equation}
in field space, where $E_1$ and $L_1$ are the right- and left-handed
selectron fields, and where the term $\sim  h_e A_e$ in the scalar
potential can give a large negative contribution. Once $A_e$ and the
corresponding soft masses at the appropriate scale (using the
corresponding RG equations) are expressed in terms of $A_0$, $M_{1/2}$
and $m_0$, the condition for the absence of such a charge and colour
breaking minimum becomes~\cite{genNMSSM2}
\beq\label{ccb}
\left(A_0-0.5\, M_{1/2}\right)^2 \lsim 9\, m_0^2+2.67\, 
M_{1/2}^2\; .
\eeq
In our analysis we will obtain relatively small values for $A_0$ ($A_0
\sim - \frac14 M_{1/2}$, cf. Fig.~\ref{fig:M12:m0.A0.tb} below), for
which eq.~(\ref{ccb}) is satisfied independently of the value of $m_0$.

More delicate could be unbounded-from-below (UFB)  directions in field
space, which are both D-flat and F-flat. In~Ref.~\cite{ccbnmssm}, it has
been clarified that such dangerous directions in the field space of the
MSSM are still present in the NMSSM, although the singlet vev $s$ gives
an additional positive contribution to the potential. Analytic
approximations to the potential along such dangerous directions have been
studied in Ref.~\cite{abelsavoy}, with the following results: 

\begin{itemize}
\vspace*{-2mm}
\item[$i)$] the inequality $m_0 \gsim 0.3\, M_{1/2}$ (for the large
$\tan\beta$ which  will be relevant here)  is an approximate  condition
for the absence of deeper minima in these directions -- this inequality
will be violated below; 
\vspace*{-2mm}

\item[$ii)$] the decay rate of the standard vacuum is usually much larger
than the age of the universe; hence we have to assume that the early
cosmology (temperature-induced positive masses squared for the squarks
and sleptons) places us into the local standard minimum of the scalar
potential. 
\end{itemize}

\subsection{Constraints  from the dark matter relic density}

In the cMSSM, small values of $m_0$ give rise to a stable charged slepton
LSP, which would be an unacceptable dark matter candidate. The slepton
LSP problem in the cMSSM with small $m_0$ can be evaded in the cNMSSM due
to the presence of the additional singlino-like neutralino which, in
large regions of the parameter space, is the true LSP \cite{hugbelpuk}. 
However, in order to be a good dark matter candidate, its relic density
should comply with the WMAP constraint \cite{wmap} 
\begin{equation}\label{wmap}
0.094 \, \lesssim \Omega_{\chi_1^0} h^2 \lesssim 0.136 \quad \quad 
(\mathrm{at \ } 2\, \sigma)\; .
\end{equation}

In the case of a singlino-like LSP, the upper bound on the relic density
implies that the singlino-like LSP mass $m_{\chi_S}$ ($=m_{\chi_1^0}$) 
has to be close to
(but somewhat below) the mass of the NLSP, which in the present case is
always the lighter (mostly right-handed) stau $\tilde\tau_1 \sim \tilde
\tau_R$: 
\beq\label{singstaumasses}
m_{\chi_S}^2 \sim m_{\tilde\tau_R}^2\,.
\eeq

Only then can the singlino co-annihilate sufficiently rapidly with the
NLSP. The condition for nearly degenerate stau and singlino masses can be
obtained by replacing in the singlino mass squared, $m_{\chi_S}^2 \sim 4
\kappa^2  s^2$~\cite{genNMSSM1,genNMSSM2,genNMSSM3,nmssmtools,slha2}, the
analytic approximation for the singlet vev as obtained from
eq.~(\ref{singpot}),
\begin{equation}\label{sapprox}
s \approx \frac{1}{4\kappa} 
\left(-A_\kappa + \sqrt{A_\kappa^2-8m_S^2}\right)\,.
\end{equation}
Noticing again that $A_\kappa \sim A_0\ (<0)$ and $m_S^2 \sim m_0^2$,
$m_{\chi_S}^2$ is approximately given by
\begin{equation}\label{singmass}
m_{\chi_S}^2 \simeq \frac{1}{2}\left(A_0^2 + \left|A_0\right|
\sqrt{A_0^2-8 m_0^2}\right) -2 m_0^2\; .
\end{equation}

An analytic approximation for the right-handed stau mass at the weak
scale, obtained by integrating the RG equations,  is given by \cite{genNMSSM2}
\beq\label{staumass}
m_{\tilde\tau_R}^2\sim m_0^2 + 0.1\, M_{1/2}^2\,.
\eeq

Inserting eqs.~(\ref{singmass}) and (\ref{staumass}) into
eq.~(\ref{singstaumasses}) and using eq.~(\ref{upperm0}), one can derive
the bound 
\begin{equation}
m_0^2 \lsim \frac{1}{15} M_{1/2}^2.
\end{equation}
In practice, however, all approximations above (notably the neglected
$\tilde \tau_R - \tilde \tau_L$ mixing) tend to overestimate $m_0$, and
the stronger bound 
\begin{equation}
m_0\lsim \frac{1}{10} M_{1/2} 
\end{equation}
holds. Hence, $m_0$ must be quite small when compared to $M_{1/2}$, and
could well vanish.

These analytic approximations allow to understand which ``hyperplane" in
the para\-meter space $M_{1/2}$, $m_0$ and $A_0$ will satisfy the WMAP
constraint of eq.~(\ref{wmap}): not only must $m_0$ be small, but $A_0$
is essentially determined  in terms of $M_{1/2}$ by
eq.~(\ref{singstaumasses}). Our numerical results (for $m_0 \sim 0$, see
below) correspond to
\beq
A_0 \sim -\frac{1}{4} M_{1/2}\; .
\eeq

Finally, as in the cMSSM~\cite{upper}, the WMAP constraint also requires
that $M_{1/2}$ should not be too large. Since the dominant annihilation
process of R-odd sparticles is now  $\tilde \tau_1 + \tilde \tau_1 \to$
SM particles, the rate decreases with $m_{\tilde \tau_1}$ (which is
roughly proportional to $M_{1/2}$), eventually becoming too small for 
$M_{1/2} \gsim $2--3 TeV. 

The routine NMSSMTools~\cite{nmssmtools}, which includes a version of the
dark matter tool \break
MicrOMEGAS~\cite{belsemenov} adapted to the NMSSM,
allows to scan the parameter space of the cNMSSM and to verify which
para\-meters $M_{1/2}$, $m_0$ and $A_0$ satisfy the WMAP constraint,
eq.~(\ref{wmap}). We recall that $\lambda$ is another free parameter of
the model, whereas $\tan\beta$ is fixed by the condition $m_S = m_0$ at
the GUT scale. In the next subsection we will discuss that LEP
constraints impose an upper bound on $\lambda$, $\lambda
\lsim 10^{-2}$, and, for illustrative purposes, we will fix $\lambda=
0.002$  throughout the remaining part of this subsection. In any case,
the following results are practically independent of $\lambda$.
\begin{center}
\begin{figure}[ht!]
\begin{center}
\begin{tabular}{cc}\hspace*{-10mm}
\psfig{file=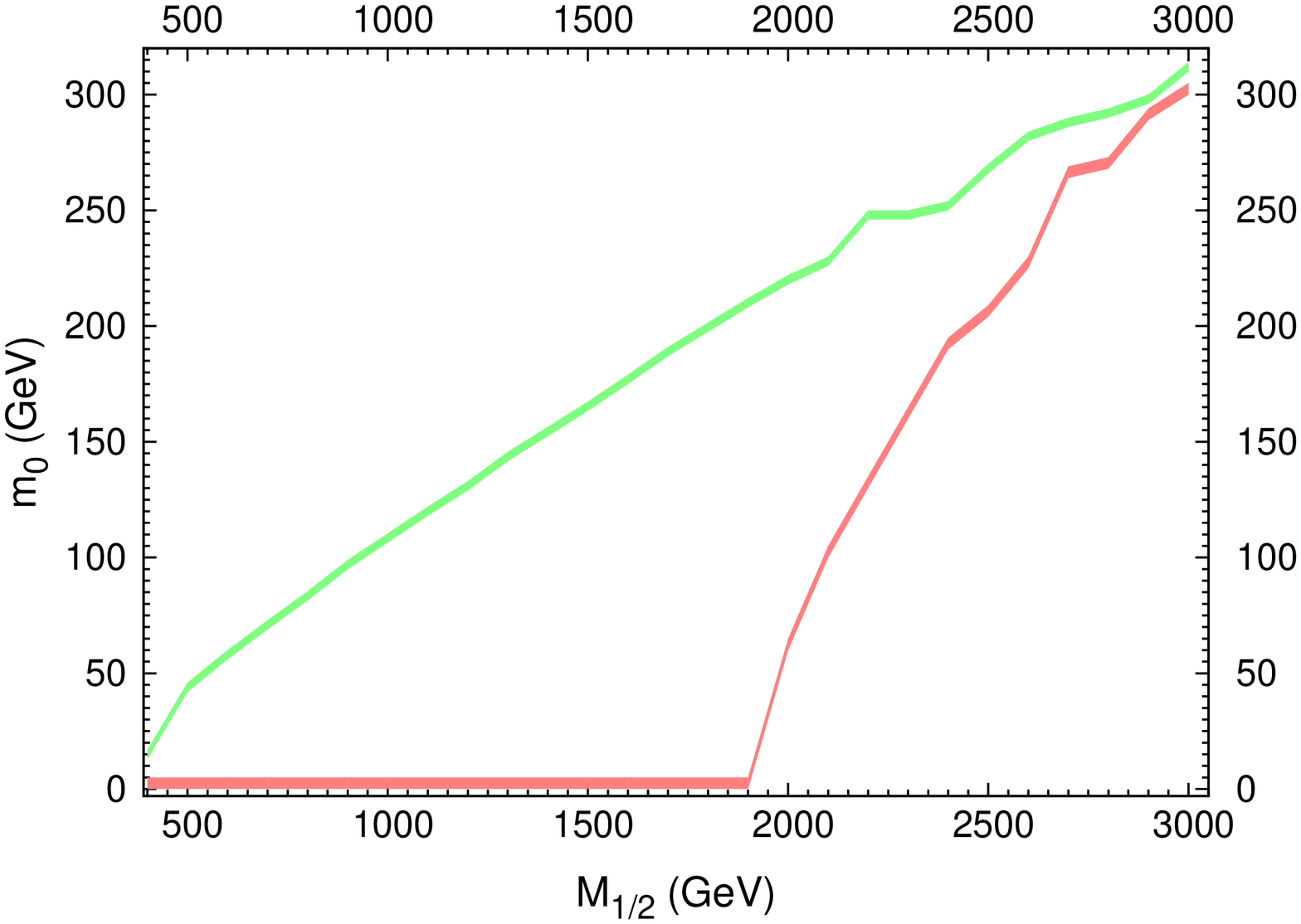, clip=, angle=270, 
width=84mm} \ \  &
\hspace*{-10mm}
\psfig{file=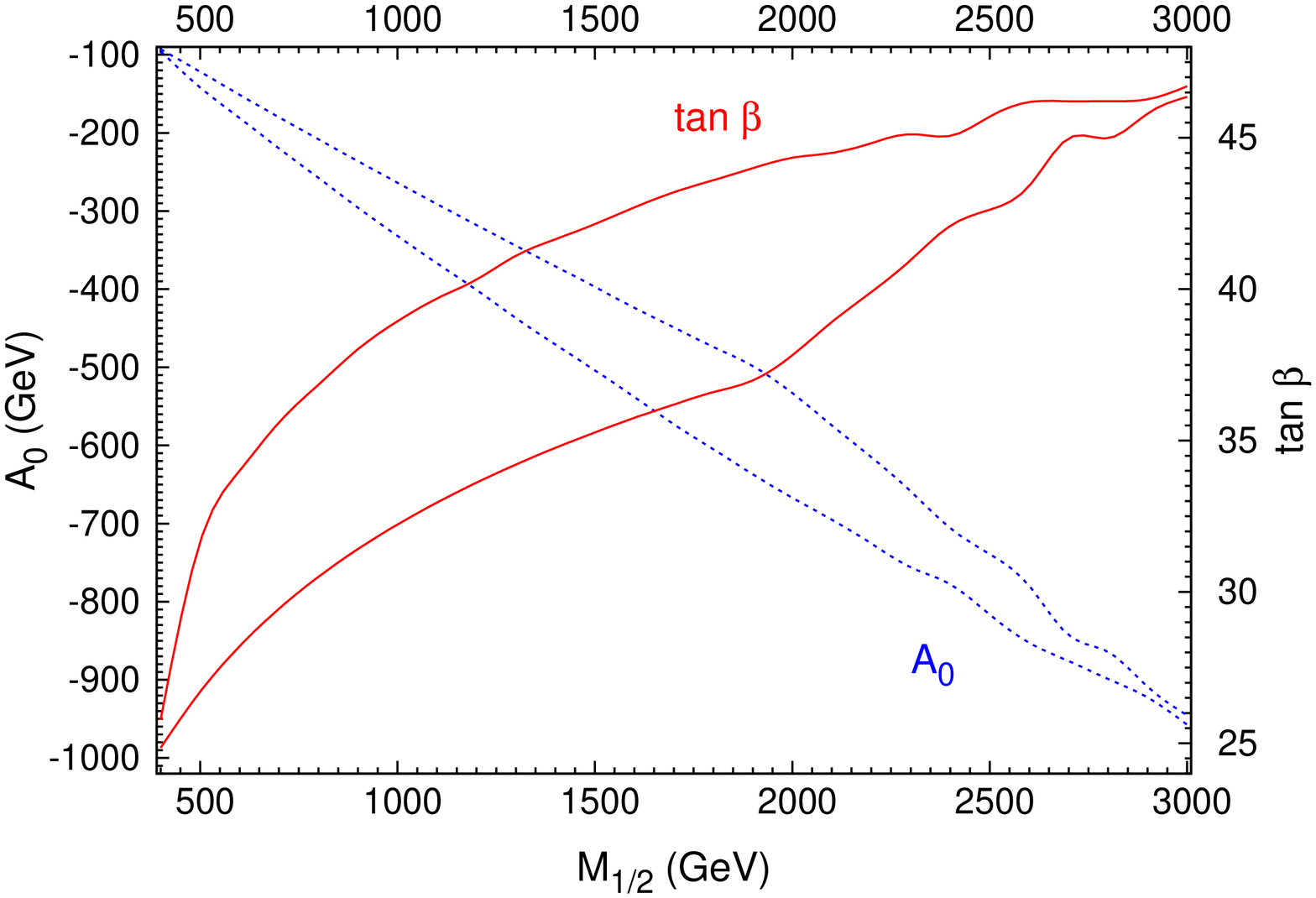, clip=, angle=270, 
width=84mm}
\end{tabular}
\caption{In the left panel, maximal and minimal values of 
  $m_0$ allowed by the WMAP constraint of eq.~(\ref{wmap}), as a function 
  of $M_{1/2}$ (in GeV). 
  In the right panel, associated range of $A_0$ (in GeV) and
  $\tan \beta$, also as a function of $M_{1/2}$. 
  In the case of $\tan \beta$ ($A_0$), the lower (upper) line corresponds
  to $m_0=0$.
}
  \label{fig:M12:m0.A0.tb}
\end{center}
\end{figure}
\end{center}

\vspace*{-5mm}
In Fig.~\ref{fig:M12:m0.A0.tb} we show the allowed range of the
parameters $m_0$ and $A_0$ as functions of $M_{1/2}$. In the left panel,
we display the contours for the minimal and maximal values of $m_0$. We
notice that $m_0=0$ is only compatible with the WMAP constraint  for
$M_{1/2}~\lsim~2$~TeV.  Moreover, as $M_{1/2}$ increases, it becomes 
increasingly difficult to satisfy the WMAP constraint and, hence, the
allowed range for $m_0$ decreases. For $M_{1/2}>3$ TeV, hardly any
parameter space survives since $\Omega_{\chi_1^0} h^2$ would be too
large. 

In the right panel of Fig.~\ref{fig:M12:m0.A0.tb}, we present the values
of $A_0$ corresponding to the maximal and minimal values of $m_0$
for a given $M_{1/2}$ (we recall that eq.~(\ref{wmap}) fixes
$A_0$ in terms of $M_{1/2}$ and $m_0$). Likewise, we display the values
of $\tan \beta$ obtained from the requirement of full scalar mass
unification  $m_S^2 \simeq m_0^2$.

We notice that $\tan\beta$ turns out to be quite large (see
Ref.~\cite{largetb} for earlier work on the NMSSM at large $\tan\beta$).
The origin of the large value of $\tan\beta$ can be understood as
follows. First, an effective $B$-parameter 
\beq
B_{\rm eff} = A_\lambda + \kappa s
\eeq 
can be defined, which plays the same r\^ole as the $B$-parameter of the
MSSM. Then, $\tan\beta$ is inversely proportional to  $\sim |B_{\rm
eff}|$. In the regime where $m_S^2 \sim m_0^2 \ll A_0^2 \sim A_\kappa^2$,
eq.~(\ref{sapprox}) gives $s \sim -A_\kappa /2 \kappa$, and thus $B_{\rm
eff}\approx \, A_\lambda -\frac{1}{2} A_\kappa$ with $A_\kappa \sim A_0$
and $A_\lambda$ determined by the RG equations. Finally, accidentally,
$A_\lambda -\frac{1}{2} A_\kappa$ happens to be small (much smaller than
$\mu_{\rm eff}$) leading to $\tan\beta \gg 1$.

The lower limit on $M_{1/2}$ of $\sim$\;400~GeV follows from the lower
bound on the lightest stau mass of {$\sim$\;100~GeV} from the negative
LEP searches \cite{pdg}. The corresponding lower limit on $M_{1/2}$
derived from eq.~(\ref{staumass}) seems somewhat weaker, but the stau
mixing has to be taken into account. For $M_{1/2} \approx 400$~GeV, we
observe from  Fig.~\ref{fig:M12:m0.A0.tb} that only $m_0 \lsim 20$~GeV is
viable. For larger values of $M_{1/2}$, values for $m_0$ up to $\sim
\frac{1}{10} M_{1/2}$ are possible (with sizable values of $\tan \beta
\gtrsim 40$). For values of $M_{1/2}$ larger than 2 TeV, compatibility
with the correct dark matter  relic density (which requires $M_{1/2} < 2$
TeV for $m_0 = 0$) can no longer be obtained for $m_0=0$: the
increasingly larger values of $\tan \beta$, as determined by the
requirement of scalar mass unification at the GUT scale,  generate a
stronger mixing in the stau sector. Then, in order to have $m_{\tilde
\tau_1} \gtrsim m_{\chi_1^0}$, a non-vanishing, albeit small, value of
$m_0$ is required. For $m_0 \neq 0$, compatibility with the WMAP bound
allows for $M_{1/2}$ up to around 3 TeV, where $m_0 \sim 300$ GeV and
$\tan\beta \sim 46$, with the upper bound on $m_0$ following from
eq.~(\ref{upperm0}).

\begin{figure}[ht!]
\begin{center}
\psfig{file=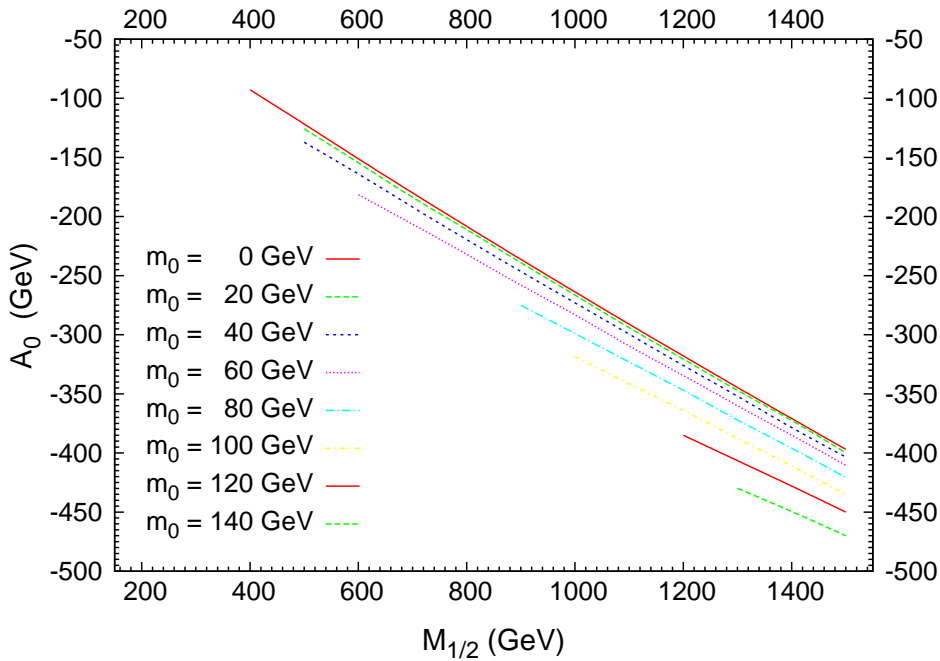, clip=, angle=270, 
width=140mm}
\caption{Distinct allowed ``lines'' in the $[M_{1/2},A_0]$ plane
compatible with the WMAP constraint of eq.~(\ref{wmap}).
From top to bottom, the lines correspond to  decreasing values of $m_0=0$
from  20 to 140 GeV.}
\label{fig:M12:A0:gen}
\end{center}
\end{figure}

Restricting ourselves to the phenomenologically more interesting regime
$M_{1/2} \lesssim$ 1.5~TeV,  we present in Fig.~\ref{fig:M12:A0:gen}
several WMAP-compatible ``lines'' in the $[M_{1/2},A_0]$ plane for fixed
values of $m_0$. Actually, the allowed 2\,$\sigma$ range for $\Omega h^2$
implies that the lines displayed correspond to ``bands'' of finite (yet
small) width. The parameter space lying below these lines is typically
excluded due to the presence of a stau LSP (and to a minor extent, also a
violation of the WMAP constraint). The upper regions delimited by each
line are excluded due to an excessively large relic density.  

\subsection{Constraints on $\lambda$ from LEP and dark matter}

Remarkably,  LEP constraints on the SM-like Higgs boson mass turn out to
be satisfied due to the relatively large stop masses and trilinear
coupling $A_{t}$ for {\em all} parameter ranges shown in 
Figs.~\ref{fig:M12:m0.A0.tb} and \ref{fig:M12:A0:gen}. However, they lead
to upper bounds on the NMSSM specific parameter~$\lambda$. For the  large
values of $\tan\beta$ obtained in this scenario,  a large value
of $\lambda$ does {\it not} lead to an increase of the SM-like Higgs
mass. On the contrary, increasing $\lambda$ simply increases the mixing
of the singlet-like CP-even scalar with doublet-like CP-even scalars. If
the singlet-like CP-even scalar mass is larger, the SM-like Higgs mass
decreases with increasing $\lambda$ and can fall below the LEP bound. If
the singlet-like CP-even scalar mass is below the SM-like Higgs mass
limit,   i.e. $\lesssim$~114~GeV, its coupling to the $Z$-boson, 
which is proportional to $\lambda$, must be sufficiently small, equally
implying an upper bound on $\lambda$.

\begin{figure}[ht!]
\begin{center}
\psfig{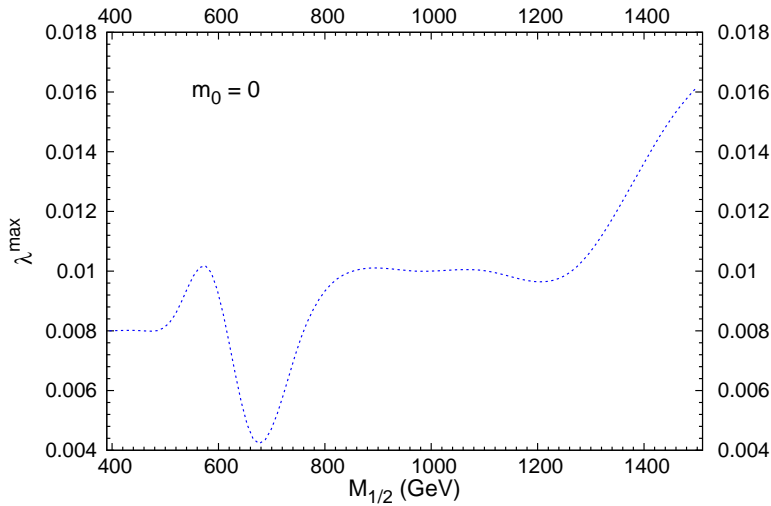}
\caption{The upper bound on  the parameter $\lambda$ as a function of 
$M_{1/2}$ (in  GeV), as obtained from LEP constraints on the NMSSM Higgs 
sector; for simplicity, we set $m_0=0$.}\label{fig:M12:A0} 
\end{center}
\end{figure}

In Fig.~\ref{fig:M12:A0}, we show the corresponding upper limits on
$\lambda$ for the case $m_0=0$. The constraint is  particularly strong in
the ``cross-over'' region near $M_{1/2} \sim 660$~GeV (see
Fig.~\ref{fig:m12:phenoH} below), where relatively small values of
$\lambda$ can generate a large mixing angle. We see that $\lambda
\lesssim 10^{-2}$ is required for all values of $M_{1/2}$; hence, the
parameter $\lambda$ will have practically no effect on the remaining
spectrum except for the singlet-like Higgs masses.

On the other hand,  a rough lower bound on $\lambda$ can be derived from
the efficiency  of singlino-like LSP annihilation in the early universe.
For the dilution of the LSP, two processes are relevant in the limit of
small $\lambda$.  The dominant annihilation process of R-odd particles is
$\mathrm{NLSP} + \mathrm{NLSP} \to X$, the NLSP being the lightest stau.
The reaction rate for this process is given by $(n_\mathrm{NLSP})^2
\sigma$, where $n_\mathrm{NLSP}$ is the corresponding time and
temperature dependent abundance, and $\sigma$ is the thermally averaged
cross section. 

The other relevant process leading to the observed relic density is
$\mathrm{LSP} + X \to \mathrm{NLSP} + X'$ (and its inverse), where $X$
and $X'$ are practically massless quarks and leptons -- this process
helps to maintain the LSP and the NLSP in thermal equilibrium. Its
reaction rate can be written as $n_\mathrm{LSP}\, n_{X}\, \sigma'$, and
depends on the small, but non-vanishing, non-singlet component of the
singlino-like LSP of ${\cal O}(\lambda/g_2)$, where $g_2$ is the SU(2)
coupling constant.  Consequently, the cross section $\sigma'$ is $\sim
\lambda^2/g_2^2 \times \sigma$, and is correspondingly suppressed for
small $\lambda$.  

Nevertheless, the process $\mathrm{LSP} + X \to \mathrm{NLSP} + X'$  is
typically faster than the annihilation process $\mathrm{NLSP} +
\mathrm{NLSP} \to X$, since near the  freeze-out temperature, the
abundance $n_X$ of quarks and leptons is  $\sim 10^9$ larger than the
abundances of the LSP and NLSP  (for $m_\mathrm{LSP} \sim
m_\mathrm{NLSP}$)~\cite{griest}. This allows to dilute the LSP density as
fast as the NLSP density. 

Only for very small $\lambda$, the reaction rate of the process
$\mathrm{LSP} + X \to \mathrm{NLSP} + X'$ can become smaller than the one
for $\mathrm{NLSP} + \mathrm{NLSP} \to X$; then the LSP will no longer be
in thermal equilibrium with the NLSP near the freeze-out temperature, but
can be considered as decoupled, implying an excessively large relic
density. According to the discussion above, this would happen for
$\lambda^2/g_2^2 \lsim 10^{-9}$ or $\lambda \lsim 10^{-5}$. Clearly,
these are rough estimates, which would merit more detailed 
investigations. In the following, we employ values of $\lambda$
sufficiently above $10^{-5}$ such that the hypothesis of thermal
equilibrium between the LSP and the NLSP near the relevant temperature
can be considered as satisfied.

We remark that, even if we hypothetically allowed for other contributions
to the dark matter relic density, this would not affect the lower bound
for $\lambda$ estimated above, nor the previous discussion on the allowed
parameter space (i.e., the derived bounds on $M_{1/2}$, $m_0$, etc.),
since all bounds originate from the {\em upper} WMAP limit on the relic
density.

\subsection{Constraints from  flavor physics}

As previously mentioned, we have checked that constraints from  $B$-meson
physics \cite{bphys} are satisfied. More precisely, agreement  within
2\,$\sigma$ between the following observables and their theoretical
values is verified by the NMSSMTools routine \cite{nmssmtools}: the 
decay branching ratios ${\rm BR}(B\to X_s\gamma)$,  ${\rm BR}(\bar{B}^+
\to \tau^+\nu_{\tau})$,  ${\rm BR}(B_s \to \mu^+\mu^-)$ and  the mass
differences $\Delta M_q$, $q=d,s$. It turns out that all regions in
parameter space investigated before (i.e. consistent with WMAP and
collider constraints) are also allowed by constraints from $B$-physics.

\begin{figure}[h!]
\vspace*{-5mm}
\begin{center}
\psfig{file=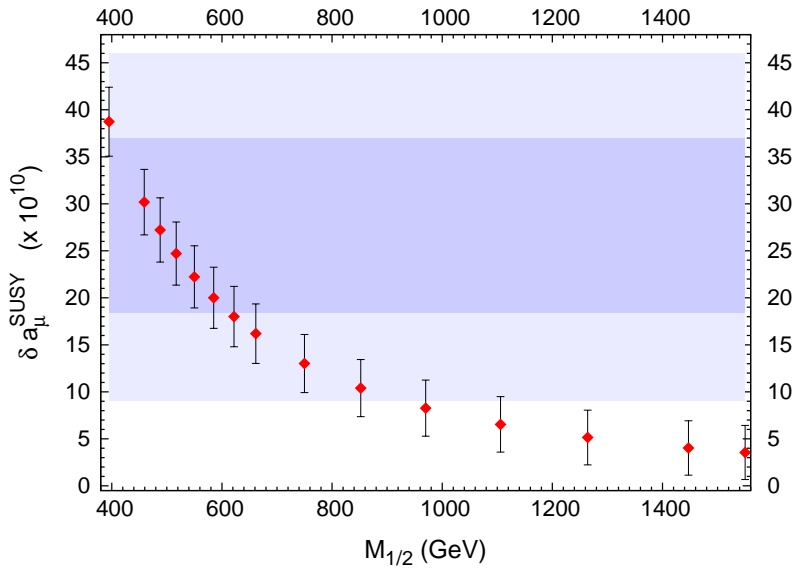, clip=, angle=270,
width=120mm}
\end{center}
\vspace*{-5mm}
\caption{$\delta a_{\mu}^{\rm SUSY}$ as a function of $M_{1/2}$ (in GeV)
  for $m_0=0$. The vertical bars denote the theoretical error, while
  darker (lighter) regions correspond to a $1\, \sigma$ ($2\, \sigma$)
  deviation from the central value of $a_\mu^{\rm exp}-a_\mu^{\rm SM}$;
  data taken from Ref.~\cite{magmu}.}
\label{magmu.f}
\end{figure}

More important constraints arise from requiring that SUSY accounts for
the $\sim 3\,\sigma$ deviation of the observed anomalous magnetic  moment
of the muon with respect to the SM expectation\footnote{A recent
measurement of the hadronic cross section $e^+ e^- \to$ hadrons  using
radiative return by the BaBar collaboration indicates that this 
discrepancy might be smaller than presently thought \cite{Davier}.}, 
$\delta a_\mu^{\rm SUSY}=(27.7 \pm 9.3)\, \times 10^{-10}$~\cite{g-2}.
The cNMSSM has been analysed in this respect in Ref.~\cite{magmu}, the
analysis being conducted for  simplicity for $m_0=0$, but this result is
practically independent of $m_0$. In Fig.~\ref{magmu.f}, we show $\delta
a_\mu^{\rm SUSY} = (g-2)_\mu^{\rm SUSY}$ as a function  of $M_{1/2}$,
depicting the 1\;$\sigma$ and 2\;$\sigma$  bands as well. 

We see that the constraint from $a_\mu$ would confine the allowed range
of $M_{1/2}$ to $M_{1/2}\lsim$ 1~TeV at the $2\,\sigma$ level, and to
400~GeV~$\lesssim M_{1/2} \lsim$ 600 - 700~GeV (where the sparticle spectrum is
not too heavy) at the $1\,\sigma$ level. In fact, the present
experimental value could be matched to arbitrarily high precision, and a
more precise measurement of $a_\mu$ would eventually lead to a prediction
of $M_{1/2}$ in the cNMSSM. In any case, in view of the desired  value for 
$\delta a_\mu^{\rm SUSY}$, the region $M_{1/2} \lsim 1$~TeV is preferred.

\section{Higgs and sparticle spectra}\label{higgs:pheno}

We now proceed to analyse in some detail the Higgs and sparticle spectra
obtained in the allowed regions of the cNMSSM parameter space, after 
imposing the bounds from Higgs and sparticle searches at LEP,
the $B$-meson constraints and the requirement of a correct cosmological
relic density as measured by WMAP for the neutralino LSP.

\subsection{The Higgs spectrum}

The Higgs sector of the (c)NMSSM contains three neutral CP-even Higgs
states $h^0_i$ ($i=1,\ 2,\ 3$), two neutral CP-odd Higgs states $a^0_i$
($i=1,\ 2$), and the charged Higgs states $h^\pm$. Their masses depend
essentially on the gaugino mass parameter $M_{1/2}$.

As in the MSSM in the decoupling regime (see Ref.~\cite{Review} for a
review), the heaviest CP-even, CP-odd and charged Higgs states form a
practically degenerate SU(2)~multiplet with a common mass above 500 GeV.
The mostly SM-like CP-even state has a mass increasing
slightly with $M_{1/2}$ from 115~GeV up to $\sim 120$~GeV.  This mass
range is only slightly above the lower limit of 114.4 GeV on the  SM
Higgs boson mass, and is compatible with the Higgs mass range favored  by
electroweak precision data as recently obtained from a global fit (in
which the central mass value is 116 GeV) \cite{Gfitter}.

The third CP-even state has a dominant singlet component; it is the only
Higgs state whose mass depends -- apart from $M_{1/2}$ -- on
$m_0$ and, to some extent, on $\lambda$: for small $M_{1/2}$ it is
lighter than the SM-like Higgs boson, escaping LEP constraints due to the
very small coupling to the $Z$~boson. For increasing values of $M_{1/2}$,
its mass increases until it becomes nearly degenerate with the SM-like
CP-even Higgs state: in this region of parameter space, which will be
subsequently denoted as the ``cross-over'' region, the singlet-like and
SM-like Higgs states strongly mix. For still larger values of $M_{1/2}$,
the mass of the singlet-like state exceeds the one of the SM-like state.

\begin{figure}[b!]
\begin{tabular}{cc}\hspace*{-6mm}
\psfig{file=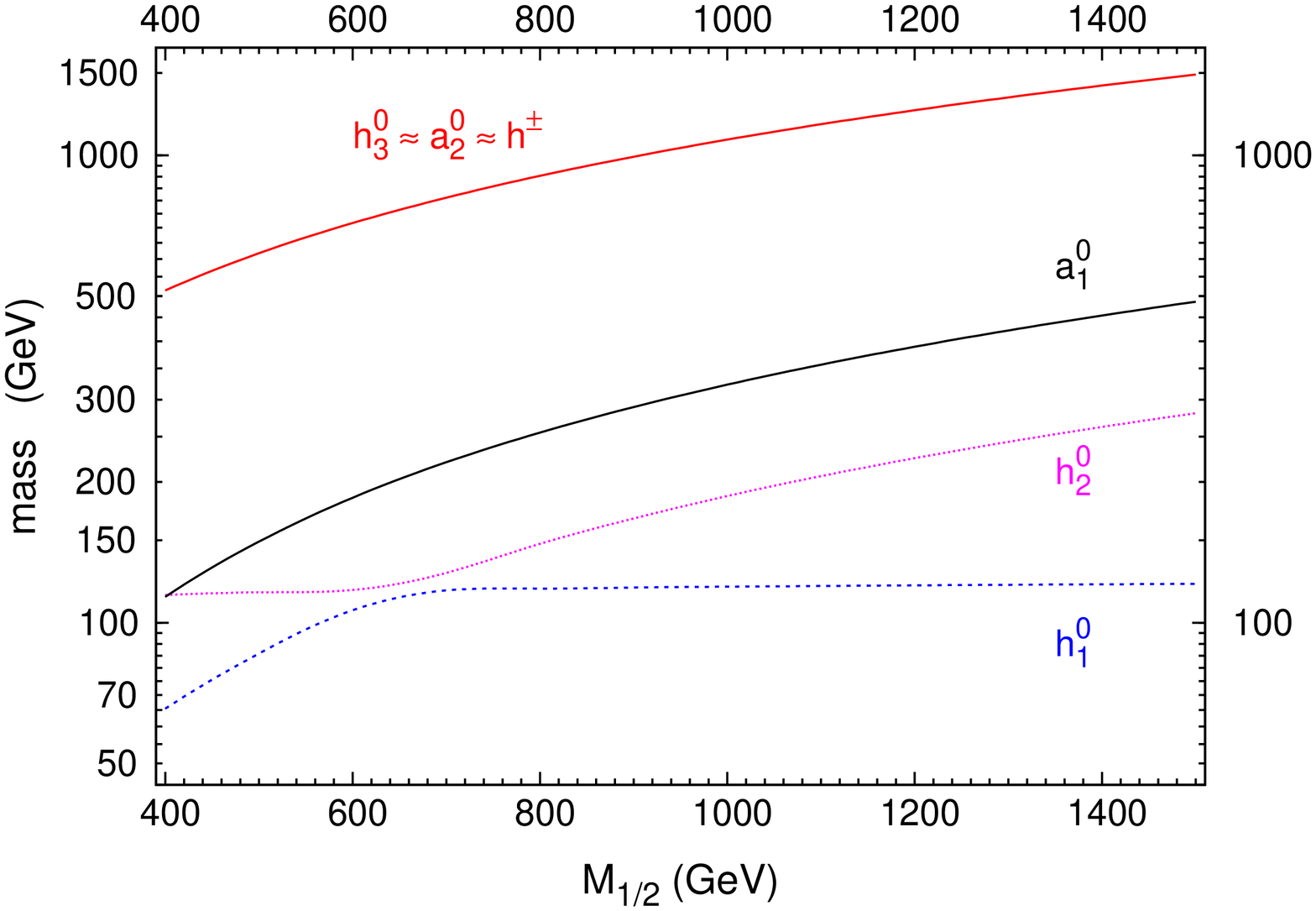, clip=, angle=270, 
width=82mm}\hspace*{-3mm}&
\hspace*{-5mm}
\psfig{file=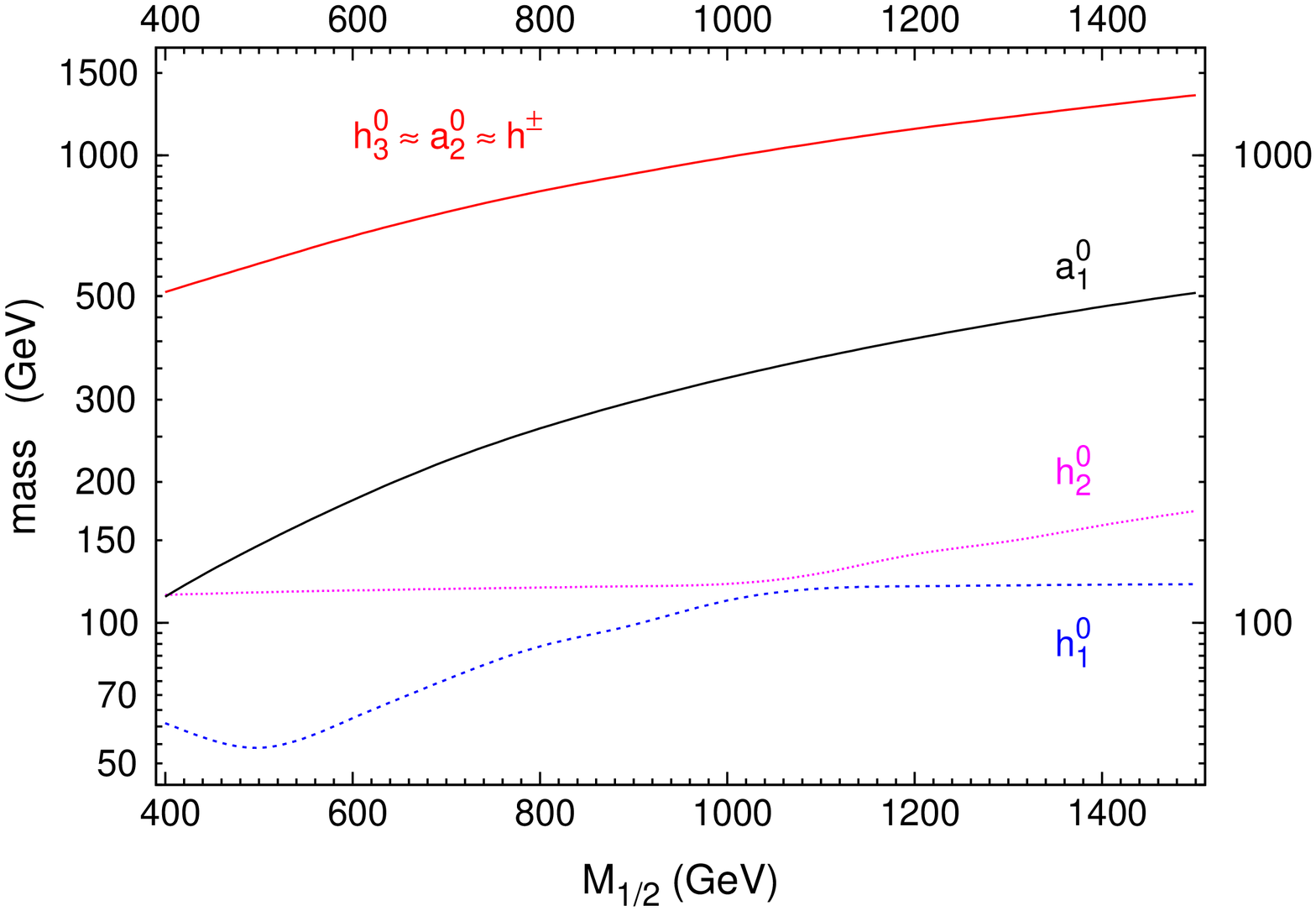, clip=, angle=270, 
width=82mm}
\end{tabular}
\caption{The Higgs masses as a function of $M_{1/2}$ (in GeV). In the
left panel we set $m_0= 0$, while in the right panel $m_0$ is  given by its
maximal value, $m_0^\mathrm{max}(M_{1/2})$. From below, the displayed
lines correspond to the states $h^0_1$ (blue/dotted), $h^0_2$
(pink/dashed),  $a^0_1$ (full/black) and $a^0_2$ (full/red) which is
degenerate with  the $h^0_3$ and $h^\pm$ states.}
\label{fig:m12:phenoH}
\end{figure}

This ``cross-over'' phenomenon is visible in Fig.~\ref{fig:m12:phenoH}, where
we display the masses of the neutral CP-even, CP-odd, and charged Higgs
bosons as a function of the parameter $M_{1/2}$. On the left-hand side,
we take $m_0=0$, while on the right-hand side we assume
$m_0=m_0^\mathrm{max}(M_{1/2})$, as given by the upper line in the left
panel of Fig.~\ref{fig:M12:m0.A0.tb}. (With the exception of the CP-odd
$a^0_1$ state, the Higgs mass spectrum is somewhat heavier in the case 
$m_0=0$.)

In Fig.~\ref{fig:m12:phenoH} we have set $\lambda = 0.002$; then, for
$m_0=0$ corresponding to the left panel, the ``cross-over'' phenomenon
occurs at $M_{1/2} \sim 660$~GeV: for $M_{1/2} \lsim 660$~GeV, the
lightest CP-even state $h_1^0$ is singlet-like, whereas the
lightest CP-even state $h_1^0$ is SM-like for $M_{1/2} \gsim 660$~GeV.
In the case $m_0= m_0^\mathrm{max}(M_{1/2})$ displayed in the right
panel of Fig.~\ref{fig:m12:phenoH}, the corresponding ``cross-over'' occurs
at $M_{1/2} \sim 1100$~GeV. Intermediate regimes for $m_0$
(with $m_0 \neq 0$) will imply a ``cross-over'' that takes place for 
$660\lesssim M_{ 1/2}\lesssim 1100$~GeV.

For the present value of $\lambda = 0.002$, the CP-even singlet mass can
be as small as 60~GeV for $m_0=m_0^\mathrm{max}(M_{1/2})$ (in the right
panel). For larger values of $\lambda$, close to the upper bound shown
in  Fig.~\ref{fig:M12:A0}, even lower $m_{h^0_1}$ values  can be
obtained, as will be discussed later.

The lighter singlet-like CP-odd scalar $a_1^0$ has a mass above
$\sim 120$~GeV (increasing with $M_{1/2}$); the heaviest CP-even and
CP-odd scalars $h_3^0$ and $a_2^0$ are practically degenerate in mass with the 
charged Higgs boson $h^\pm$, with masses above $\sim 520$ GeV.

\subsection{A possible explanation of  the excess in Higgs searches at
LEP}

In addition to the $1.7\,\sigma$ signal for a SM--like CP-even Higgs particle 
with a mass close to 115~GeV,  the combined results on Higgs searches of
the four LEP experiments via the Higgs-strahlung process $e^+ e^- \to h\,
Z$, followed by the Higgs decay $h \to b \bar{b}$, show a $2.3\,\sigma$
excess of events corresponding to a Higgs mass around 
98~GeV~\cite{Schael:2006cr,LEP98excess}. The number of excess events
amounts to about 10\% of those expected for a SM Higgs boson $h^{\rm SM}$
with the same mass. It can be explained either

\begin{itemize}
\vspace*{-2mm}

\item[$i)$] by a reduced coupling of a candidate Higgs boson to the SM
gauge bosons, $C_{h}^V = g_{h Z Z}/g_{h^{\rm SM} Z Z} \approx
\mathcal{O}(\sqrt{0.1})$, 
\vspace*{-2mm}

\item[$ii)$] by a reduced branching ratio of a candidate Higgs boson into
$b \bar{b}$ final states.
\end{itemize}

In the unconstrained MSSM, the two excesses can be explained
\cite{MSSM-excess} by the  presence of a SM-like Higgs boson with a mass
of $\approx 115$~GeV  (the heavier CP-even $H$ state), while the lighter
CP-even state $h$ has a mass close to $\approx 98$ GeV and reduced
couplings to the $Z$ boson. 

Within the context of the unconstrained NMSSM, the explanation $ii)$ of
the excess for Higgs masses around  98~GeV has been proposed
~\cite{Dermisek:2005gg}. In this case, one can have a CP-even Higgs boson
with a corresponding mass  and SM-like $ZZh$ couplings, i.e. $C_{h}^V
\sim \mathcal{O}(1)$, but a reduced branching ratio into $b \bar{b}$
final states due to a dominant decay into pairs of very light CP-odd
bosons  $h_1^0 \to a_1^0\, a_1^0$ with $m_{a_1^0} < 2 m_b$. Then, the light
pseudoscalar $a_1^0$ can only decay into $\tau^+ \tau^-$ and eventually 
light quark and gluon pairs, rendering it compatible with corresponding
searches at LEP~\cite{Schael:2006cr}. 

In the cNMSSM, the parameter space somewhat below the cross-over
regions contains neutral Higgs scalars with masses $\sim$ 100~GeV and
couplings $C_{h_1^0}^V \approx \mathcal{O}(\sqrt{0.1})$. As an example,
for 560~GeV $\lesssim  M_{1/2} \lesssim $~575~GeV (with $m_0=0$ and
$\lambda = 0.005$), one finds for the masses of the two lighter CP-even
Higgs states 97~GeV $\lesssim m_{h_1^0} \lesssim $ 101~GeV and
$m_{h_2^0} \approx 117$ GeV. In this case, the reduced coupling of the
lightest scalar Higgs boson to SM gauge bosons would lie in the  range
$0.28 \lesssim |C_{h_1^0}^V| \lesssim 0.33$ for $h_1^0$, so that the
cNMSSM could indeed account for the observed 2.3\,$\sigma$ excess around
$m_h \approx 98$ GeV  at LEP. In addition, since the mass of the nearly
SM--like $h_2^0$ state is $m_{h_2^0} \approx 117$ GeV with $|C_{h_2^0}^V
\sim 0.9|$, and in view of the error of a few GeV in the determination
of the radiative corrections to the Higgs masses (expected to be, as in
the MSSM, of the order of 3 GeV, see e.g.  Ref.~\cite{Mh-error}), the
1.7\,$\sigma$ excess at a Higgs mass $\approx 115$ GeV could be
explained as well. 

Thus, in the cNMSSM, the region in parameter space corresponding to 
small $M_{1/2}$ can describe not
only the deviation of the $(g-2)_\mu$ from the SM expectation, 
but both excesses of Higgs--like events at
LEP as well.

\subsection{The sparticle spectrum}

\begin{figure}[b!]
\vspace*{-5mm}
\begin{tabular}{cc}\hspace*{-6mm}
\psfig{file=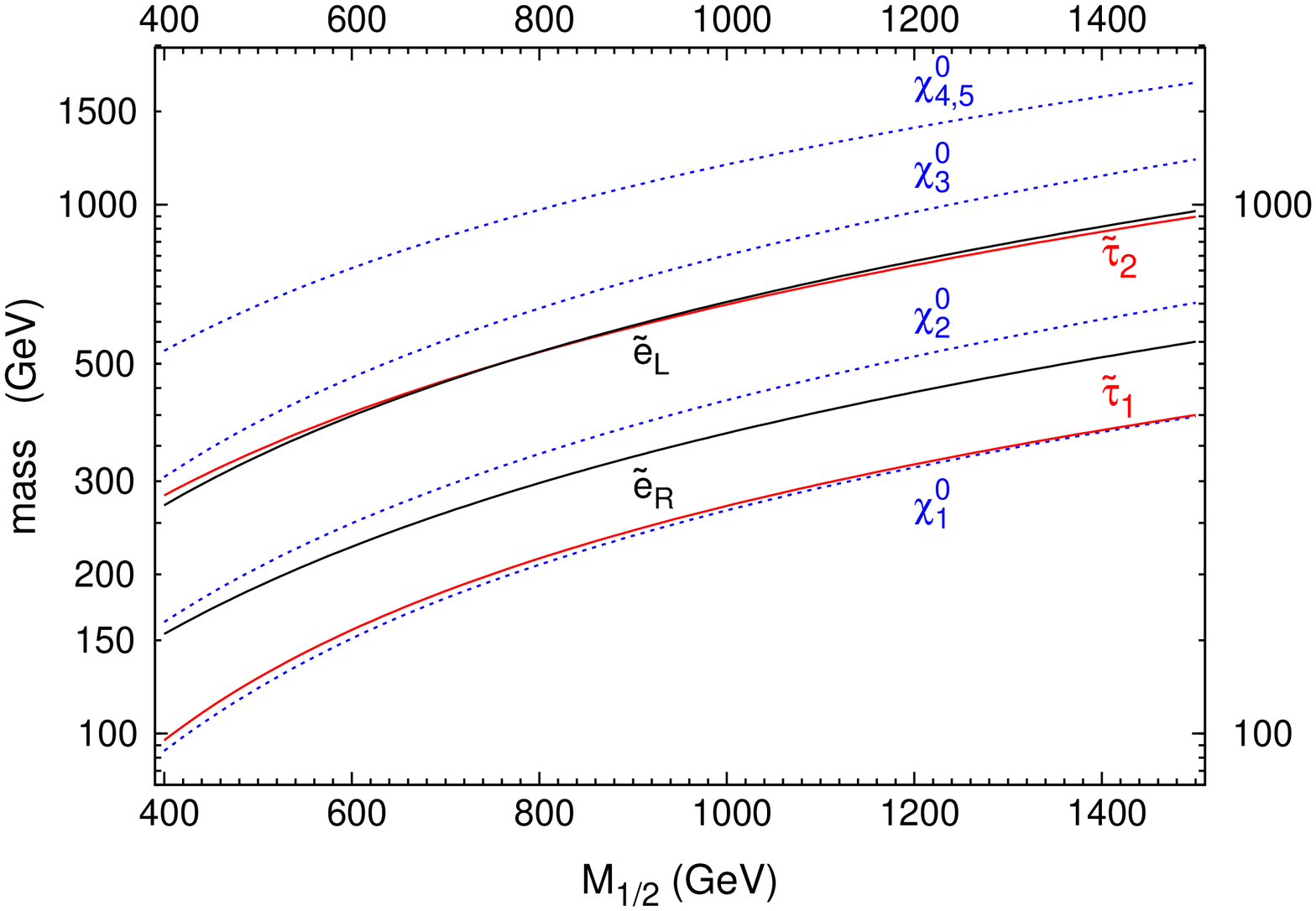, clip=, angle=270, 
width=83mm} \hspace*{-5mm}&
\hspace*{-5mm}
\psfig{file=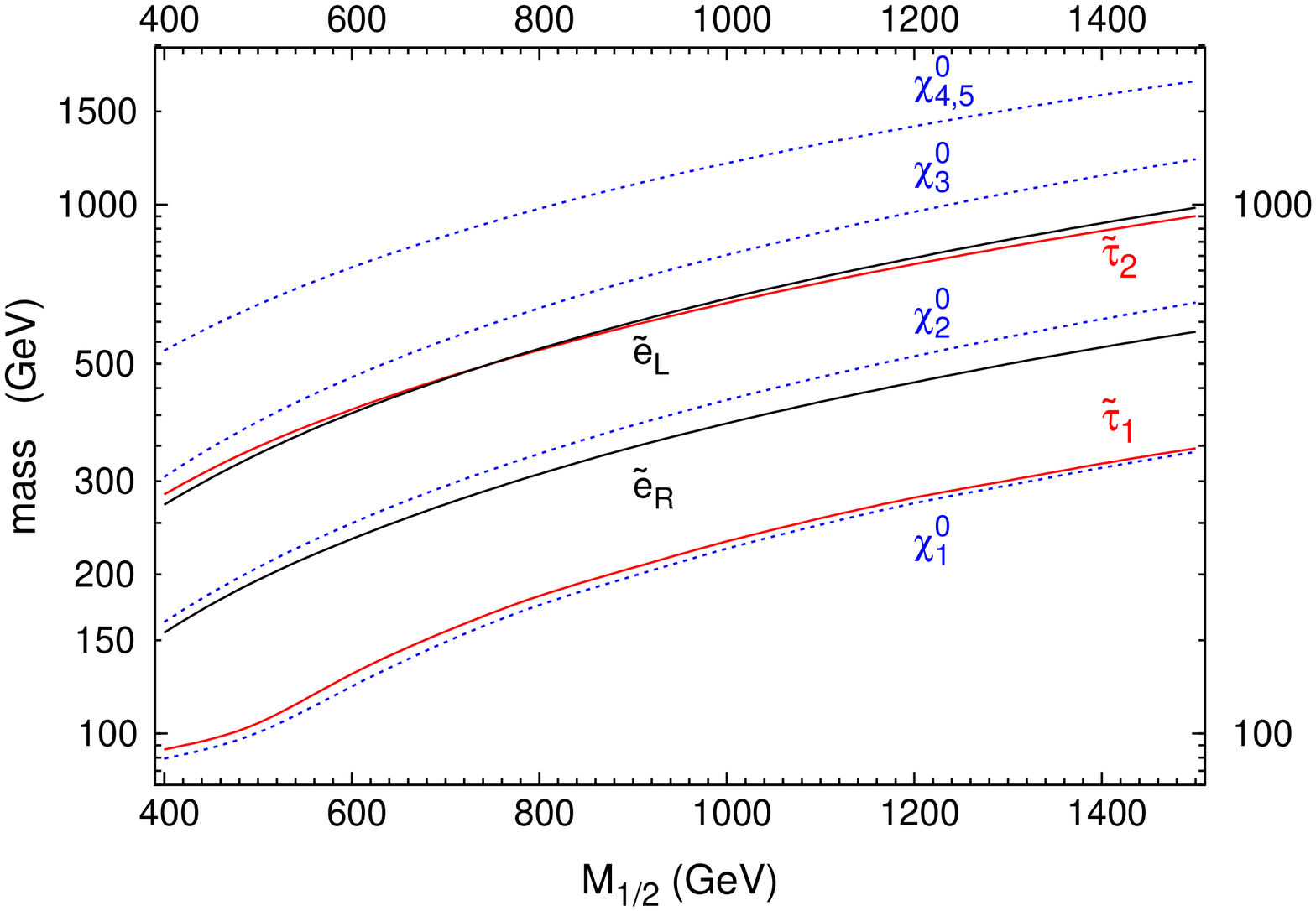, clip=, angle=270, 
width=83mm}
\end{tabular}
\caption{Neutralino (blue/dotted lines), selectron (black/full lines) 
  and stau (red/full lines) masses 
  as a function of $M_{1/2}$ (in GeV); on the left-hand side  
  we set $m_0=0$, while 
  on the right-hand side  we set $m_0=m_0^\mathrm{max}(M_{1/2})$.
  In both panels the states are ordered in
  mass as $m_{\chi^0_1} \lesssim m_{\tilde \tau_1} <
  m_{\tilde e_R} < m_{\chi^0_2} < m_{\tilde \tau_2}
  \lesssim m_{\tilde e_L} < m_{\chi^0_3}
  < m_{\chi^0_{4,5}}$.  
 The charginos
$\chi_1^\pm$ and $\chi_2^\pm$ are degenerate in mass with,
respectively, $\chi_3^0$ and $\chi_{4,5}^0$. 
  }\label{fig:m12:phenoX0st}
\end{figure}

Let us now turn to the sparticle spectrum, starting with the neutralino and
slepton mass spectra shown in Fig.~\ref{fig:m12:phenoX0st}. As for the Higgs
bosons, we display the case $m_0=0$ in the
left-hand panel, and $m_0=m_0^\mathrm{max}(M_{1/2})$ in the
right-hand panel. 

The two nearly degenerate sets of lower lines in both panels 
correspond to the masses
of  the $\chi_1^0$ singlino-like LSP (blue/dotted) and the lighter stau
$\widetilde{\tau}_1$ NLSP (red/full). The mass difference between these
two states is smaller than $\sim 8$~GeV, as required in order to
obtain a cosmological relic density for the singlino $\chi_1^0$ 
compatible with WMAP. 

The pattern for the masses of the charginos and the heavier neutralinos
(blue/dotted lines) follows the one of the MSSM, once the proper relabeling
of the states is made. Since the low-energy value of the higgsino mass
parameter $\mu_{\rm eff}$ is generally quite large, $\mu_{\rm eff}\gsim 
M_2$, the heavier  neutralino states $\chi_{4}^0$ and $\chi_{5}^0$ are
higgsino-like with masses $\sim \mu_{\rm eff}$. 
The states $\chi_2^0$ and $\chi_3^0$  are, respectively,
bino and wino-like with masses $m_{\chi_3^0} \approx 2 m_{\chi_2^0}
\approx M_2$ (with $M_2\approx 0.75 M_{1/2}$). 
The charginos $\chi_1^\pm$ and $\chi_2^\pm$ are nearly degenerate in
mass with, respectively, the wino-like $\chi_3^0$ and  the higgsino-like
$\chi_{4,5}^0$ states. 

For completeness, we have also depicted the masses of the left and right
handed selectrons, $\tilde e_L$ and $\tilde e_R$ (black/full lines), 
which are  degenerate in mass with, respectively, the  smuons $\tilde
\mu_L$ and $\tilde \mu_R$. The right-handed states are always lighter
than the left-handed ones and the mass pattern is such that  $m_{\tilde
e_L} < m_{\chi_2^0} < m_{\tilde e_R}$. Of course, all these states are
heavier than the NLSP $\tilde \tau_1$ (in fact, $\tilde e_L$ is
almost  degenerate with $\tilde \tau_2$). Due to SU(2)
symmetry, the sneutrinos  have approximately the same masses as the
left-handed sleptons. 

\begin{figure}[t!]\begin{center}
\vspace*{-5mm}
\psfig{file=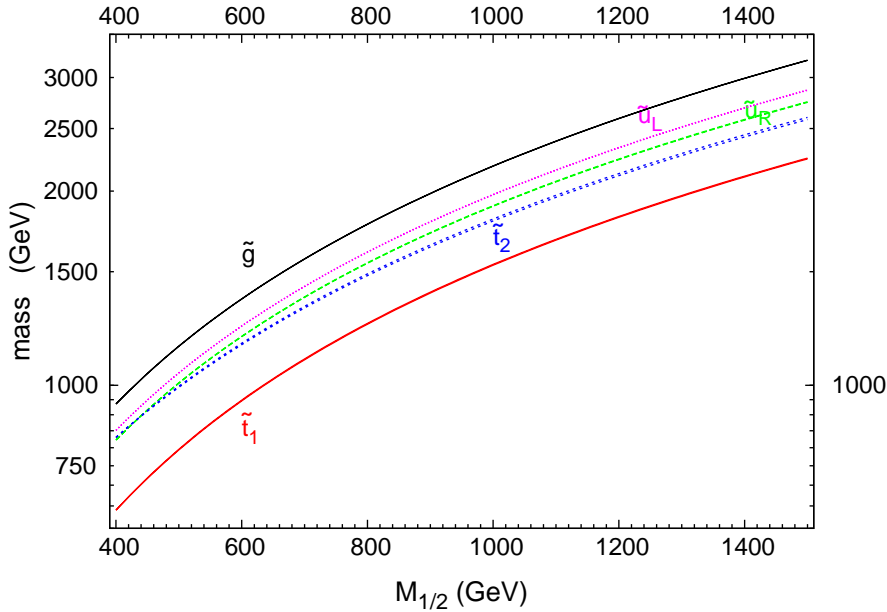, clip=, angle=270,
width=130mm}
\caption{Squark and gluino masses as a function of $M_{1/2}$
  within the range obtained by taking $m_0=0$ or 
  $m_0=m_0^\mathrm{max}(M_{1/2})$. The states are mass ordered as
  $m_{\tilde t_1}\! <\! m_{\tilde t_2} \!<\! m_{\tilde u_R}\! < \!
  m_{\tilde u_L} \!< \!m_{\tilde g}$.}
  \label{fig:m12:phenoGQL}
\end{center}
\vspace*{-5mm}
\end{figure}

In Fig.~\ref{fig:m12:phenoGQL}, we display the masses of the gluino and 
of up-type squarks as a function of $M_{1/2}$. Here, the
variation of the parameter $m_0$ in the range $0 \lesssim m_0 \lesssim
m_0^\mathrm{max}(M_{1/2})$ results only in a hardly visible ``width'' of
the lines. The masses of the $c$-quarks are degenerate with those of the
$u$-squarks and have not been displayed in  the figure.  

We first note that the gluino is always heavier than the squarks, a
consequence of the small ratio $m_0/M_{1/2}$; this feature has
important consequences as will be discussed later. The
left- and right-handed up-type squarks are almost degenerate
(the mass difference being less than 5\%), and also nearly degenerate
with the first and second  generation down-type squarks, which are 
not shown in Fig.~\ref{fig:m12:phenoGQL}. The top squarks
are lighter than the $u$-squarks, the mass of the lighter stop $\tilde
t_1$ being $\approx 20\%$ smaller. Thus, $\tilde t_1$ is the lightest
strongly interacting particle. The bottom squarks have masses somewhat
below the heavier stop $\tilde t_2$.  

Actually, since the singlet sector practically decouples, most of the
spectrum could also be obtained in the cMSSM, provided the present values
of  $M_{1/2}$, $m_0$ and $A_0$ are used. We have
explicitly verified that the obtained spectra of the
non-singlet states for the  two points of the cNMSSM  parameters
discussed in the next subsection agree with those obtained in the MSSM
with the program Suspect \cite{Suspect} (up to small differences
due to the treatment of the higher order radiative corrections).

\subsection{Examples of spectra}

The most relevant features of the sparticle and Higgs spectrum can be
represented by two points P1 and P2, for which the distinct cNMSSM
inputs, as well as the spectra, are summarized in the left part of
Table~\ref{tab:b1b2}.  These two points illustrate the low
and intermediate  $M_{1/2}$ regime in the cNMSSM. 

\begin{table}[!ht]
\begin{center}
\begin{tabular}{ll}
\begin{minipage}{7cm} 
\begin{tabular}{||l||c||c||}
\hline
\hline
& \phantom{eV}P1\phantom{eV}
&\phantom{eV}P2\phantom{eV}\\
\hline
$M_{1/2}$ (GeV) & 500  &    1000  \\
\hline
$m_0$ (GeV) & 0  & 0  \\ 
\hline
$A_0$ \ \ \ (GeV) &-122  & -263 \\
\hline
$\tan \beta$ & 26.7 &   32.2\\
\hline\hline
$\mu_\mathrm{eff}$ \ \ \  (GeV)& 640 & 1185 \\
\hline
$M_2$ \ \ \  (GeV)&  390& 790  \\ \hline\hline
$m_{h_1^0}$ \ \ \ (GeV) &  86&  119\\
\hline
$m_{h_2^0}$\ \ \  (GeV) &  116&    187\\
\hline
$m_{h_3^0}$ \ \ \ (GeV) &  610&    1073\\
\hline
$m_{a_1^0}$\ \ \  (GeV) &  149&   323\\
\hline
$m_{\chi_1^0}$\ \ \  (GeV) &  122&  264 \\
\hline
$m_{\chi_2^0}$ \ \ \ (GeV) &  206&   427\\
\hline
$m_{\chi_3^0}$\ \ \  (GeV) &  388&   802\\
\hline
$m_{\chi_{4,5}^0}$\ \ \  (GeV)  &645   &  1190\\
\hline
$m_{\chi^\pm_1}$\ \ \  (GeV)  &  388&  801\\
\hline
$m_{\chi^\pm_2}$\ \ \  (GeV)  &  658 &  1198\\
\hline
$m_{\tilde g}$ \ \ \ (GeV)  &  1150&    2187\\
\hline
$m_{\tilde u_L}$\ \ \  (GeV)  &  1044&  1973 \\
\hline
$m_{\tilde u_R}$\ \ \  (GeV)  &  1007&   1895\\
\hline
$m_{\tilde t_1}$ \ \ \ (GeV)  &  795&  1539\\
\hline
$m_{\tilde t_2}$\ \ \  (GeV)  &  997&   1810\\
\hline
$m_{\tilde b_1}$\ \ \  (GeV)  &  931&   1760\\
\hline
$m_{\tilde b_2}$\ \ \  (GeV)  &  983&    1817\\
\hline
$m_{\tilde e_L}$\ \ \  (GeV)  &  334&   654\\
\hline
$m_{\tilde e_R}$\ \ \  (GeV)  &  190&  370\\
\hline
$m_{\tilde \nu_l}$\ \ \  (GeV)  &  325&  650 \\
\hline
$m_{\tilde \tau_1}$\ \ \  (GeV)  &  127&  269\\
\hline
$m_{\tilde \tau_2}$\ \ \  (GeV)  &  343&    647\\
\hline
$m_{\tilde \nu_\tau}$\ \ \  (GeV)  &  318&  631  \\
\hline
\hline
\end{tabular}
\end{minipage}
& \quad
\begin{minipage}{6cm} \vspace*{-9.1cm}
\begin{tabular}{||l||c||c||}
\hline
\hline
& \phantom{eV}P1$^\prime$\phantom{eV}
&\phantom{eV}P2$^\prime$\phantom{eV}\\
\hline
$M_{1/2}$ (GeV) &  500 & 1000 \\
\hline
$m_0$ (GeV) &  40 & 107 \\ 
\hline
$A_0$ \ \ \ (GeV)   &  -137&  -327\\
\hline
$\tan \beta$ &  30.2&  38.4\\
\hline\hline
$\mu_\mathrm{eff}$ \ \ \  (GeV)&  642   &  1192\\
\hline
$M_2$ \ \ \  (GeV)& 390 & 791  \\ \hline\hline
$m_{h_1^0}$ \ \ \ (GeV) &    64&   116\\
\hline
$m_{h_2^0}$\ \ \  (GeV) &    116 &   127\\
\hline
$m_{h_3^0}$ \ \ \ (GeV) &   588&   989\\
\hline
$m_{a_1^0}$\ \ \  (GeV) &   149&    333\\
\hline
$m_{\chi_1^0}$\ \ \  (GeV) &   107&   226\\
\hline
$m_{\tilde \tau_1}$\ \ \  (GeV)  &   112&    235\\
\hline
\hline
\end{tabular}
\end{minipage}
\end{tabular}
\end{center}
\vspace*{-3mm}
\caption{Input parameters and low-energy spectra for four points of the 
cNMSSM with two distinct $M_{1/2}$ regimes. On the left are Points P1 and
P2 with $m_0=0$, $M_{1/2}=500$ GeV and 1 TeV. On the right 
P1$^\prime$ and P2$^\prime$ with $m_0 \neq 0$, for which we only display 
those masses  whose values vary with $m_0$.
In all 
cases, we have set $\lambda=0.002$. }\label{tab:b1b2} 
\vspace*{-5mm}
\end{table}

In the right part of Table \ref{tab:b1b2}, we display the
resulting spectrum for the Higgs, LSP and NLSP states for
$m_0 \neq 0$.  While the input $M_{1/2}$ is the same,  the input
values of  $A_0$ and $\tan\beta$ have been slightly adjusted  in order to
obtain acceptable values for the dark matter density and the unification of
the singlet mass. In the points P1 and P1', the CP-even $h_1^0$ state is
singlet-like, with a somewhat lower value of $m_{h_1^0}$ for P1'.  The
$h_1^0$ state is SM-like for P2, while  $h_1^0$ and $h_2^0$ have a very
similar non-singlet components in P2' implying similar couplings to gauge
bosons and quarks.

\section{Prospects for collider searches}

\subsection{Sparticle and Higgs decays}

The most interesting aspects of the spectrum of the cNMSSM -- as compared
to the cMSSM -- are the presence of a singlino-like LSP with a mass just
below the one of the stau NLSP, and the fact that all squarks are lighter
than the gluino. The singlino-like LSP with its small coupling to all
other sparticles will strongly modify the sparticle decay chains, since
now all sparticles will decay via the stau NLSP. Also, squark and gluino
decay chains will differ from most MSSM-like scenarios. These two features
will have important consequences for sparticle searches at the LHC.

\subsubsection{Sparticle decay branching ratios}

The branching ratios for the gluinos, squarks, sleptons and
charginos/neu\-tralinos are shown in Table~\ref{tab:P1P3:decay} for the
points P1 and P2 with $m_0=0$ (the situation being similar in the primed
points with $m_0 \neq 0$). They have been obtained by applying the 
program SUSYhit~\cite{SUSYhit}, which calculates the decay widths and
the branching ratios of the Higgs and SUSY particles of the MSSM, to
the  NMSSM with a practically decoupled singlet sector. A few comments
are in order. \smallskip

\begin{table}[ht!]
\begin{center}
\begin{tabular}{ll}
\begin{minipage}{7cm} 
\begin{tabular}{||l||c|c||} \hline \hline
\ \ BR (\%) & \phantom{eV}P1\phantom{eV}& \phantom{eV}P2\phantom{eV}\\
\hline\hline
$\phantom{l_j}\tilde g \to {\tilde q}_L \,\bar q$ & 17.7  & 14.4
\\ \hline
$\phantom{l_j}\tilde g \to \tilde q_R \,\bar q$ & 33.6  & 27.5
\\ \hline
$\phantom{l_j}\tilde g \to \tilde b_1 \,\bar b$ & 16.5  & 12.8
\\ \hline
$\phantom{l_j}\tilde g \to \tilde b_2 \,\bar b$ &  10.9 & 10.3
\\ \hline
$\phantom{l_j}\tilde g \to \tilde t_1 \,\bar t$ & 21.2  & 22.4
\\ \hline
$\phantom{l_j}\tilde g \to \tilde t_2 \,\bar t$ & --  & 12.5
\\ \hline\hline
$\phantom{l_j}\tilde q_L \to \chi_3^0 \,q$ & 31.7  & 32.3
\\ \hline
$\phantom{l_j}\tilde q_L \to \chi_1^\pm \,q^\prime$ & 62.7  & 64.3
\\ \hline
$\phantom{l_j}\tilde q_R \to \chi_2^0 \,q$ &  99.7 & 99.9
\\ \hline
$\phantom{l_j}\tilde l_L \to \chi_2^0 \,l$ &  100 &  100
\\ \hline
$\phantom{l_j}\tilde l_R \to l \,\tilde \tau_1 \,\tau$ &  $\gtrsim$ 99 & 
$\gtrsim$ 99
\\ \hline
$\phantom{l_j}\tilde \nu_l \to \chi_2^0 \,\nu_l  $ &  100&100 
\\ \hline
$\phantom{l_j}\tilde \nu_\tau\to \chi_2^0 \,\nu_\tau$ & 13.8 & 6.8
\\ \hline
$\phantom{l_j}\tilde \nu_\tau \to \tilde \tau_1 \,W$  & 86.2 &93.2 
\\ \hline\hline
\end{tabular}
\end{minipage}
&
\begin{minipage}{7cm} 
\begin{tabular}{||l||c|c||} \hline \hline
\ \ BR (\%) & \phantom{eV}P1\phantom{eV}& \phantom{eV}P2\phantom{eV}\\
\hline\hline
$\phantom{l_j}\chi_2^0 \to \tilde \tau_1 \,\tau$ &  88.3 & 74.3
\\ \hline
$\phantom{l_j}\chi_2^0 \to \tilde l_R \, l$ &  11.7 & 25.7
\\ \hline
$\phantom{l_j}\chi_3^0 \to \tilde l_L \,l$ & 22.1  & 28.4
\\ \hline
$\phantom{l_j}\chi_3^0 \to \tilde \nu_l \,\nu_l$ & 27.1  & 29.2
\\ \hline
$\phantom{l_j}\chi_3^0 \to \tilde \tau_1\, \tau$ & 24.9  & 8.8
\\ \hline
$\phantom{l_j}\chi_3^0 \to \tilde \tau_2\, \tau$ & 6.9  & 14.8
\\ \hline
$\phantom{l_j}\chi_3^0 \to \tilde \nu_\tau\, \nu_\tau$ &  16.9 & 18.3
\\ \hline\hline
$\phantom{l_j}\chi_1^\pm \to \tilde \nu_l \,l$ & 29.3  & 29.9
\\ \hline
$\phantom{l_j}\chi_1^\pm \to \tilde l \,\nu_l$ & 20.8  & 27.8
\\ \hline
$\phantom{l_j}\chi_1^\pm \to \tilde \nu_\tau \,\tau$ & 18.4  & 18.9
\\ \hline
$\phantom{l_j}\chi_1^\pm \to \tilde \tau_1 \,\nu_\tau$ &  24 & 8.7
\\ \hline
$\phantom{l_j}\chi_1^\pm \to \tilde \tau_2 \,\nu_\tau$ &  -- & 14.3
\\ \hline \hline
\end{tabular}
\end{minipage}
\end{tabular}
\end{center}
\caption{Dominant decay modes of the squark, slepton, gluino, neutralino
and chargino states for the two points P1 and P2, for which the spectrum 
is given in Table~\ref{tab:b1b2} (and where $\lambda=0.002$). 
They have been obtained using SUSYhit~\cite{SUSYhit}; $q$ and $l$ denote 
first and second generation quarks and leptons, respectively.}
\label{tab:P1P3:decay}
\end{table}

-- As the gluino $\tilde g$ is heavier than all squarks, it can decay
via two--body decays into all quark-squark pairs. The
branching ratio into $t \,\tilde t_1$ final states is somewhat larger
($\sim 20\%$) as a consequence of the larger phase space due to the
lighter $\tilde t_1$ states. \smallskip

-- All squarks (including the stops) 
decay into neutralinos or charginos plus the corresponding
quark. For right-handed squarks $\tilde q_R$ of the first two generations,
the branching ratio of the decay $\tilde q_R \to \chi_2^0\, q$ is nearly
100\%, the $\chi_2^0$ state being dominantly bino-like. In the case of the
left-handed $\tilde q_L$ states, the branching ratios are $\sim \frac13$
and $\sim \frac23$ for the decays into the neutral
$\tilde q_L \to q \chi_3^0$  and charged $\tilde q_L \to q' \chi_1^-$
wino states, respectively.\smallskip

-- Regarding the decays of the electroweak gauginos, the preferred decay 
channel of the state $\chi_2^0$ is  $\chi_2^0\, \to\, \tilde \tau_1 \,
\tau$ which has a branching ratio of $\sim 90\%$ (for P1) 
as a result of the more
favorable phase space;  the remaining $\sim 10\%$ are the decays
$\chi_2^0\, \to\, \tilde l_R \, l$, where $l = e^\pm$ or $\mu^\pm$. 
The wino-like $\chi_3^0$ and $\chi_1^\pm$ decay $\sim 50\%$
into first/second generation slepton+lepton states and $\sim
50\%$ into third generation $\tilde \tau \tau, \tilde \nu_\tau \nu_\tau$
states; the reason for the breaking of  lepton universality is again the
more favorable phase space.\smallskip   

-- Finally, while the left--handed first/second generation sleptons
$\tilde l_L$ decay to 100\% into $l \chi_2^0$ final states, the
right-handed sleptons $\tilde l_R$ essentially decay via the three--body
channel  $\tilde l_R\, \to\, l\, \tilde\tau_1 \, \tau$; this decay mode
has also been discussed  in Ref.~\cite{3body}, albeit in a different
context. The branching
ratio for the two-body decay mode $\tilde l_R\, \to \chi_1^0\, l$ is
well below the percent level; 
the decay into the bino $\chi_2^0$ and a lepton is 
forbidden by phase space. The reason for the dominance of the three-body
decay is that the two--body decay can occur only via the bino-component
of ${\cal O}(\lambda)$ of the mostly singlino-like $\chi_1^0$, and
hence is extremely small, even for the maximally possible values of $\lambda
\sim 0.01$ shown in Fig.~\ref{fig:M12:A0}.  On the other hand, the
three--body decay occurs through the virtual exchange of the bino-like
$\chi_2^0$, whose virtuality is not very large as the $\tilde l_R$ and
$\chi_2^0$ masses are comparable. \smallskip

Hence, practically {\em all} sparticle decay chains contain the
$\tilde\tau_1$ NLSP. The $\tilde\tau_1$ life time can be very large, but
it will finally decay into the singlino-like LSP and a tau lepton,
$\tilde \tau_1 \, \to\,\chi_1^0 \,\tau$.

\subsubsection{Displaced vertices}

For very small $\lambda$, the couplings between the 
$\tilde \tau_1, \chi_1^0$ and $\tau$ states might
be sufficiently small, resulting into a stau track of
$\mathcal{O}(\mathrm{few \,mm})$
that  might be visible \cite{cascades,singdecay}. Hence, displaced
vertices at high-energy colliders such as the LHC  from long-lived staus could
be a ``smoking gun'' signature of the cNMSSM. Here we present some details of
the computation and the resulting possible track lengths.

Following Ref.~\cite{bartl}, the partial width of the stau decay into the
lightest neutralino and tau lepton can be written as
\begin{equation}\label{staudec1}
\Gamma(\tilde \tau_1 \to \chi_1^0 \tau)\, = \,
\frac{\rho^{1/2}(m^2_{\tilde \tau_1},m^2_{\tau} ,m^2_{\chi_1^0})}{16
  \pi m^3_{\tilde \tau_1}} 
\, [(a_{11}^2 + b_{11}^2)\,(m^2_{\tilde \tau_1}-m^2_{\tau} -
m^2_{\chi_1^0}) - 4 a_{11}b_{11} m_{\tau}m_{\chi_1^0}]\,,
\end{equation}
where  $\rho$ is the phase-space function, $\rho(x,y,z)=x^2+y^2+z^2-2 x y-
2 xz - 2 y z$, and $a_{11}$, $b_{11}$ are the $\chi_1^0\,\tilde \tau_1
\,\tau_{R,L}$ couplings, defined as
\begin{align}
a_{11} \,= \,& -g_2 \sqrt{2}\, \tan \theta_\mathrm{W}\, N_{11}\, \sin
\theta_\tau - h_{\tau} \,N_{13}\,\cos \theta_\tau\,,\nonumber \\
b_{11} \,= \,&\frac{g_2}{\sqrt{2}}\,(N_{12}+\tan \theta_\mathrm{W}\,
N_{11})
\,\cos \theta_\tau\,- h_{\tau} \,N_{13}\,\sin\theta_\tau\,.
\end{align}
$h_{\tau}$ and $\theta_\tau$ denote the tau Yukawa coupling and stau
mixing angle,  $g_2$ the SU(2) coupling constant  and $N_{1i}$ denote
the composition of the lightest neutralino, $\tilde \chi_1^0 = N_{11}
\tilde B+ N_{12} \tilde W + N_{13} \tilde H_d^0 + N_{14} \tilde H_u^0 +
N_{15} \tilde S$. In the regime of small $\lambda$,  the bino
($N_{11}$), wino ($N_{12}$) and higgsino ($N_{13}$ and $N_{14}$) 
components of the mostly singlino-like neutralino $\chi_1^0$ are 
all proportional 
to $\lambda$~\cite{cascades}. Hence it is convenient to introduce nearly
$\lambda$-independent coefficients $\alpha$ and $\beta$ defined as
$a_{11}^2 + b_{11}^2 = \lambda^2\,\alpha  $ and  $2a_{11} b_{11} =
\lambda^2\,\beta$.

In the relevant limits $\Delta m \equiv m_{\tilde \tau_1} -
m_{\chi_1^0} \ll m_{\tilde \tau_1} \sim m_{\chi_1^0}$ and 
$m_\tau \ll m_{\tilde \tau_1} \sim m_{\chi_1^0}$, the expression 
eq.~(\ref{staudec1}) for the stau decay width can be simplified 
as~\cite{cascades}
\begin{equation}\label{gamstau}
\Gamma(\tilde \tau_1 \to \chi_1^0 \tau)\, \approx \,
\lambda^2\,\frac{\sqrt{\Delta m^2 - m_\tau^2}}{4 \pi m_{\tilde \tau_1}} 
\,(\alpha \Delta m - \beta m_\tau)\; ,
\end{equation}
which summarizes the essential dependence of the stau decay width on 
$\lambda$ and $\Delta m$.

\begin{figure}[t!]\begin{center}
\begin{tabular}{cc}\hspace*{-7mm}
\psfig{file=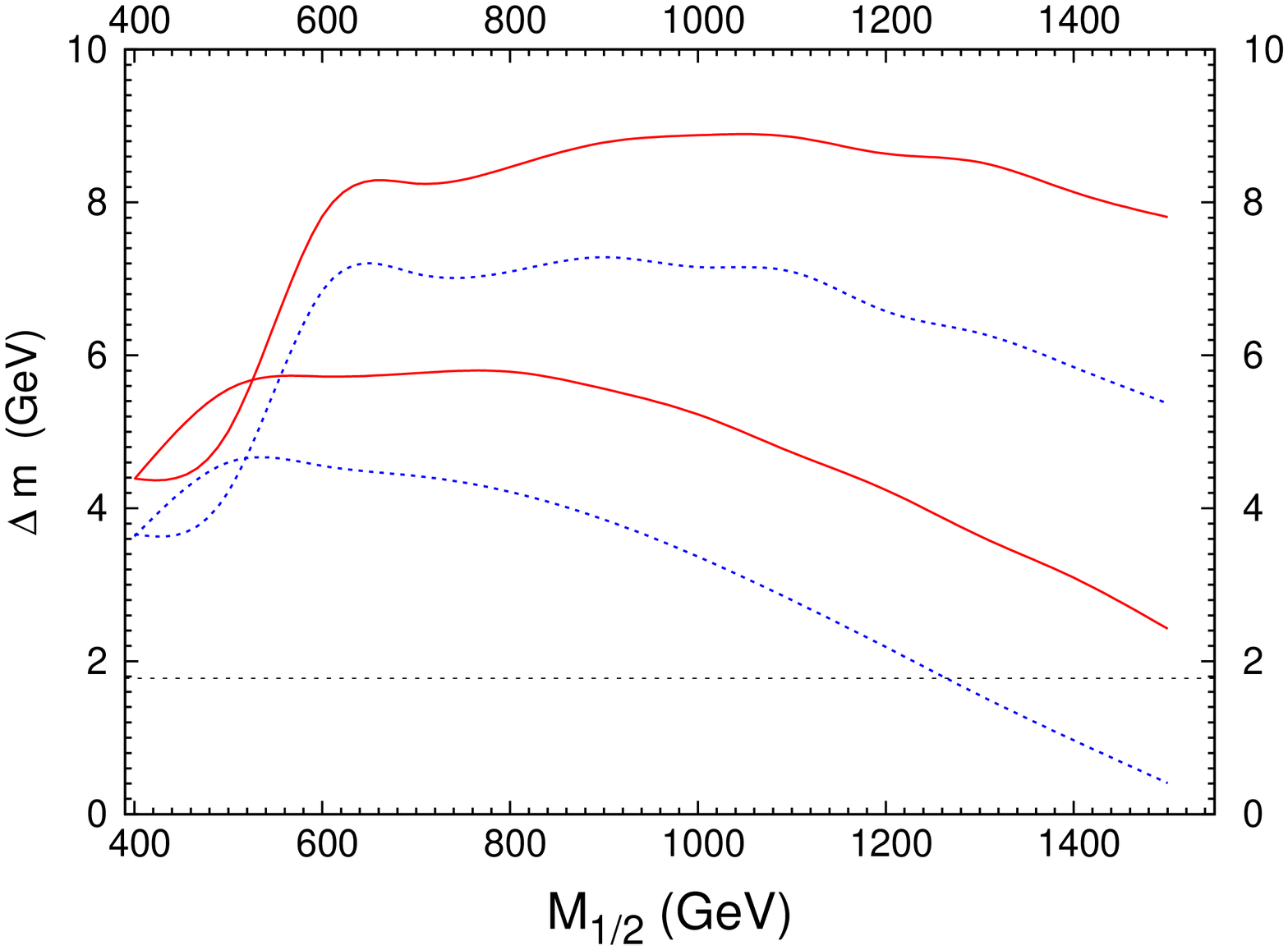, clip=,
angle=270, width=90mm}  &\hspace*{-15mm}
\psfig{file=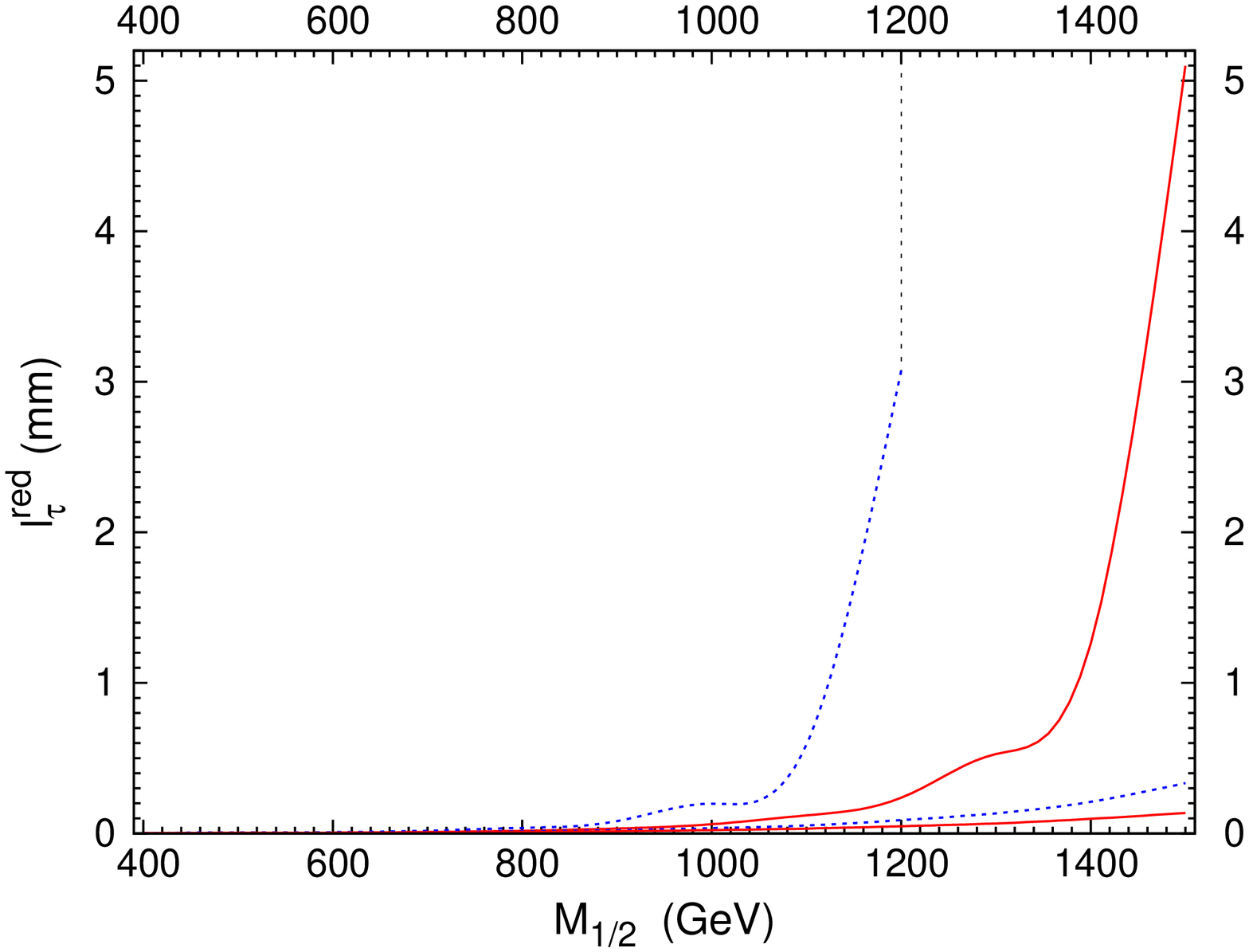, clip=, angle=270, 
width=90mm}
\end{tabular}
\caption{On the left panel, the 
  range of neutralino-stau mass differences $\Delta m$ which are
  allowed by the $2\,\sigma$ error bars of the WMAP constraint,
 as a function of $M_{1/2}$. The lower lines
  correspond
  to $m_0=0$, the upper ones to its maximal possible value. In each
  pair, full (red) lines denote the maximal $\Delta m$ -- associated
  $\Omega_{\chi_1^0} h^2 |_\mathrm{max}$, while dotted (blue) correspond
  to $\Omega_{\chi_1^0} h^2 |_\mathrm{min}$. The horizontal line denotes
  $m_\tau$.
  On the right panel, the
   maximal and minimal ``reduced'' stau lengths of flight, 
  $l^\mathrm{red}_{\tilde  \tau_1}$ (in mm), as a function of $M_{1/2}$,
  for $\lambda=10^{-3}$, for the neutralino-stau mass differences given
  in the left panel.
  }\label{fig:m12:alphabeta} 
\end{center}
\vspace*{-3mm}
\end{figure}

The coefficients $\alpha$ and $\beta$ still depend somewhat on
$M_{1/2}$ and $m_0$, and their numerical values smoothly decrease with
$M_{1/2}$; for 400 GeV $\lesssim M_{1/2} \lesssim $ 1500 GeV, one has
0.01 $\gsim \alpha \sim \beta \gsim$ 0.0001.
The mass splitting $\Delta m$ does not only depend on $M_{1/2}$ and $m_0$: 
the allowed $2\,\sigma$ range for the relic density 
in eq.~(\ref{wmap}) allows $\Delta m$ to vary within a
small window at fixed $M_{1/2}$ and $m_0$, but slightly fluctuating
$A_0$. In the left panel of Fig.~\ref{fig:m12:alphabeta}, we show the
corresponding ranges for $\Delta m$. Note that for $\Delta m < m_\tau$
the decay of the $\tilde \tau_1$ has to proceed via a virtual $\tau$,
implying a tiny partial width. 

On the right panel of Fig.~\ref{fig:m12:alphabeta} we plot the reduced stau 
length of flight,
$l^\mathrm{red}_{\tilde \tau_1}\,=\,  
\hslash\,c/\Gamma(\tilde \tau_1 \to \chi_1^0\tau)$, 
as a function of $M_{1/2}$ for $\lambda = 10^{-3}$. For other values of
$\lambda$, the reduced length of flight can be obtained by rescaling
$l^\mathrm{red}_{\tilde \tau_1}$ by a factor $(10^{-3}/\lambda)^2$.
For $M_{1/2} \gtrsim 1200$ GeV and $m_0=0$, the lifetime of the stau can
be extremely large as a consequence of the NLSP-LSP mass difference
approaching the $m_\tau$ threshold, which corresponds to the vertical
dotted line that extrapolates the upper-most curve in the right panel of
Fig.~\ref{fig:m12:alphabeta}.

The stau length of flight in the laboratory frame is given by 
\begin{equation} 
l_{\tilde \tau_1}\,=\,l^\mathrm{red}_{\tilde  \tau_1}\,
\sqrt{\beta_{\tilde  \tau_1}^2/(1-\beta_{\tilde  \tau_1}^2)}\,,
\end{equation}
where $\beta_{\tilde  \tau_1}= v_{\tilde \tau_1}/c$ is the $\tilde \tau_1$ 
velocity. A realistic
estimate of $l_{\tilde \tau_1}$ requires the knowledge of
$\beta_{\tilde  \tau_1}$  and hence of the production processes of the
lightest stau. 

A hint on realistic
values for $\beta_{\tilde  \tau_1}$ can be obtained from the
gauge-mediated SUSY breaking ATLAS benchmark point GMSB5, where it is
advocated that over 99\% of the staus (which will decay into a gravitino
LSP) have $\beta_{\tilde  \tau_1} \gtrsim 0.7$ \cite{gmsb:stau}.

\begin{figure}[!ht]\begin{center}
\begin{tabular}{cc}\hspace*{-5mm}
\psfig{file=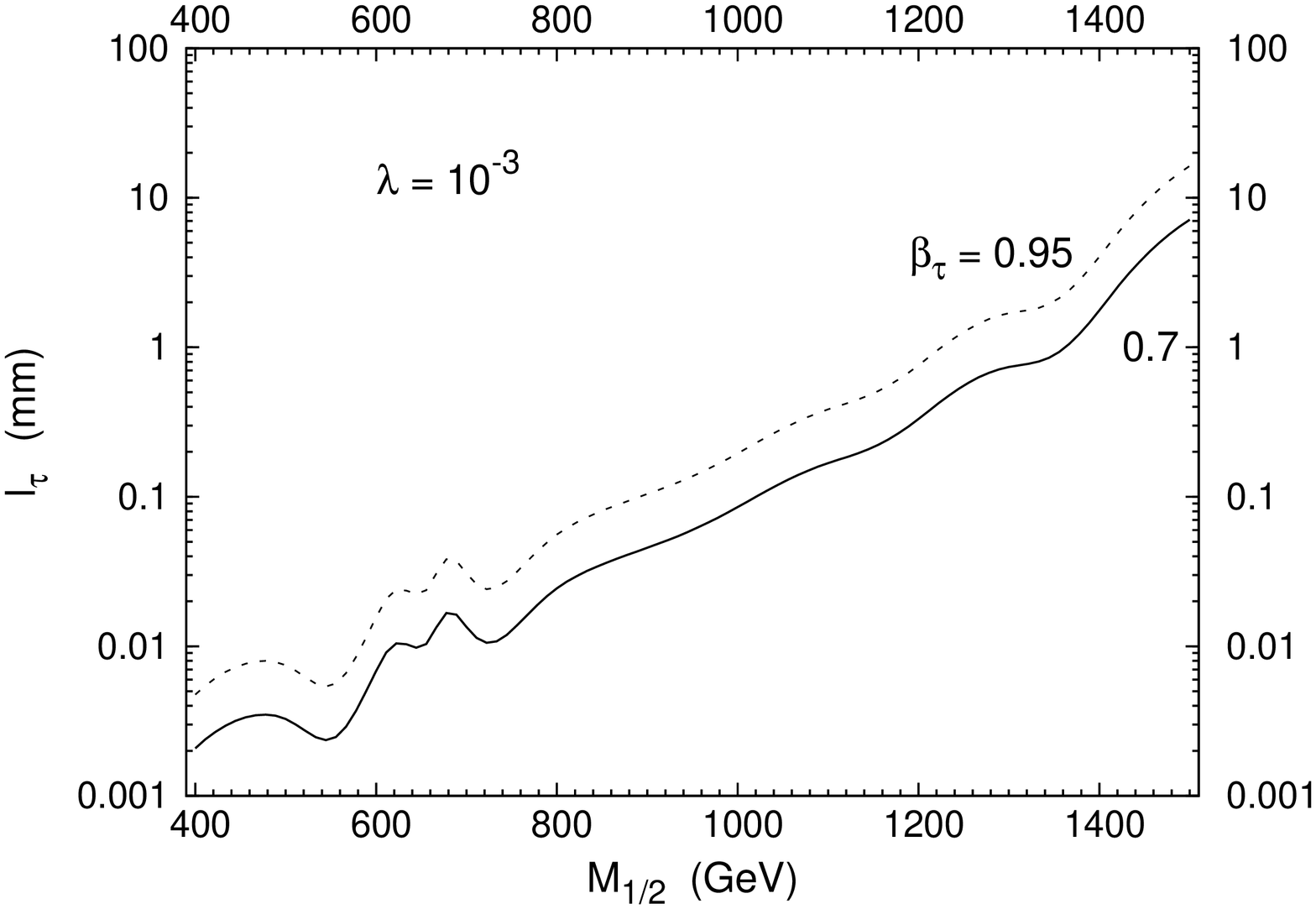, clip=, angle=270,
width=83mm} & \hspace*{-12mm}
\psfig{file=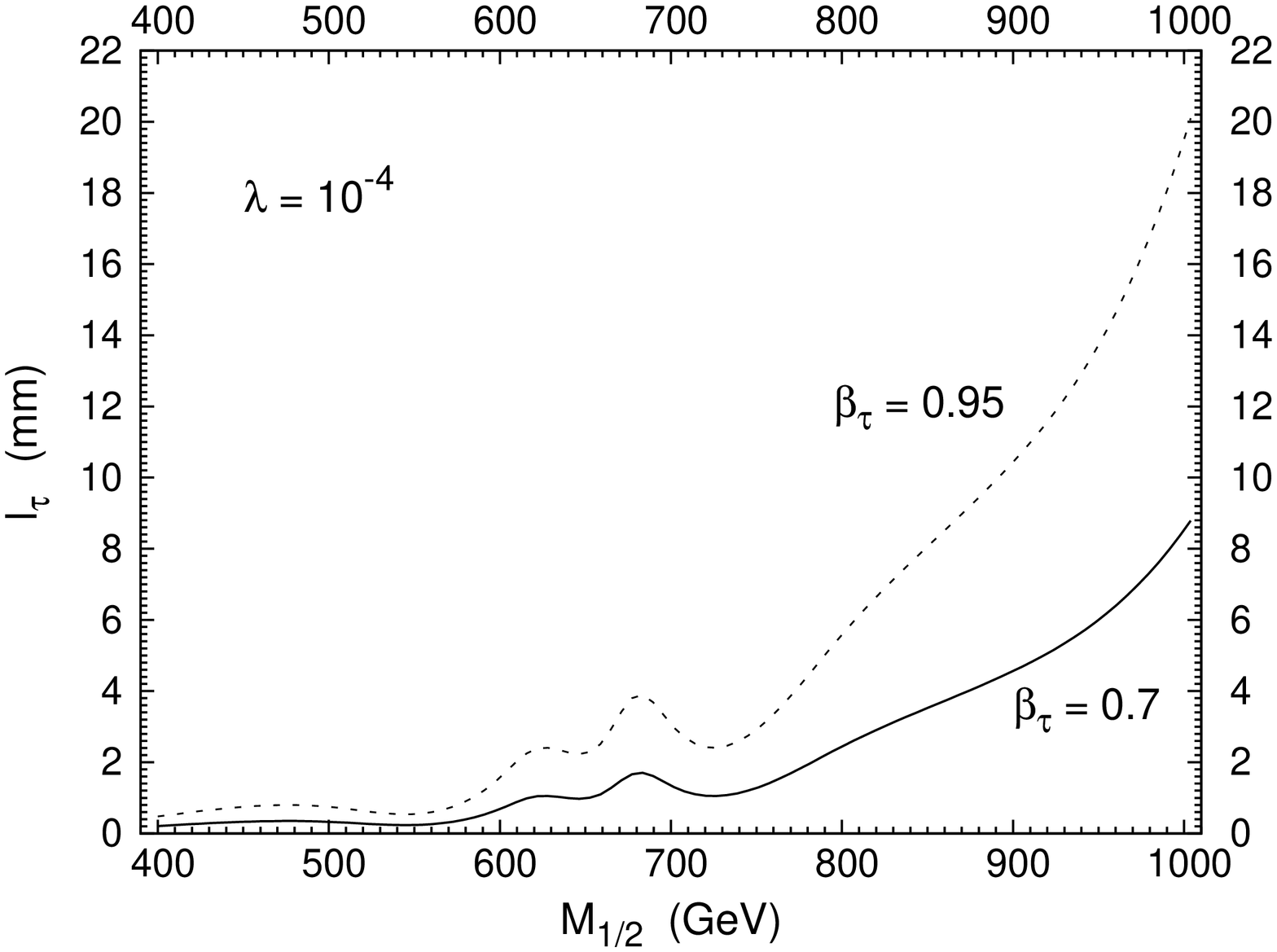, clip=, angle=270,
width=83mm} \end{tabular}
\caption{The stau length of flight $l_{\tilde  \tau_1}$ (in mm), as a function 
  of $M_{1/2}$ (in GeV), for $m_0=0$  and $\lambda=10^{-3}$
  (left panel) and $\lambda=10^{-4}$ (right panel). In both panels,   
   the upper line
  corresponds to $\beta_{\tilde  \tau_1}=0.95$, while for the lower one we
  have chosen $\beta_{\tilde \tau_1}=0.7$. Note the different scales of
  the $M_{1/2}$-axis.}
\label{fig:betal}
\end{center}
\vspace*{-6mm}
\end{figure}

In Fig.~\ref{fig:betal}, we display  the decay length $l_{\tilde  \tau_1}$
as a function of $M_{1/2}$ for the choice $m_0=0$, taking an
intermediate value of $\Delta m$ within the range presented on the left
panel of Fig.~\ref{fig:m12:alphabeta}. We consider two values of
$\beta_{\tilde  \tau_1}$, and two distinct regimes for $\lambda$. As
mentioned above, even for $\lambda$ close to its upper limit (left panel
of Fig.~\ref{fig:betal}), visible lengths of flight  $l_{\tilde  \tau_1}$,
of $\mathcal{O}({\rm mm})$, are possible for the lightest cNMSSM stau
for large $M_{1/2} \sim 1.4$ TeV.

On the right panel of Fig.~\ref{fig:betal}  we consider a smaller value
$\lambda \sim 10^{-4}$, and focus on the possibilities within the range
of $M_{1/2} < 1$~TeV favored by the data on $(g-2)_\mu$, where one
can still obtain a stau length of flight as large as a few centimeters.
For larger values of $M_{1/2}$ and $\lambda \lsim 10^{-4}$, $l_{\tilde 
\tau_1}$ can become as large as $\mathcal{O}({\rm 10 \,cm})$ and even
$\mathcal{O}(1{\rm \,m})$.  

Long-lived stau NLSPs are also possible in MSSM-like models, if the
gravitino is the true LSP. However, at least if $m_{3/2}$ is not much
smaller than $M_\mathrm{SUSY}$, the stau length of flight will always
exceed the size of the detector (see \cite{deroeck} for recent analyses
of collider signatures of such scenarios). For gauge-mediated SUSY
breaking models, where $m_{3/2}\ll M_\mathrm{SUSY}$, implications of
long-lived stau NLSPs for SUSY searches at high energy colliders such as
the LHC have been discussed for example Ref.~\cite{gmsb:stau}.  For the
NMSSM, no such studies exist at present. 

In any case,
although visibly displaced vertices from long-lived stau decays
are not an unavoidable prediction within the cNMSSM, they are a very
interesting possibility of the parameter space associated to small
$\lambda$ and/or $\Delta m$.

\subsubsection{Higgs decays}

In the cNMSSM, the decay pattern of the non singlet Higgs particles will
follow that of the MSSM in the decoupling regime \cite{Review}.  

The lightest non-singlet like CP-even Higgs boson is
SM-like with a mass between 115 and 120~GeV, implying that it will
dominantly decay into $b\bar b$ pairs with a branching ratio larger
than $\approx 70\%$, followed by decays into $c\bar c, \tau^+\tau^-$
and $gg$ pairs, with branching ratios of the order of 5\%. The branching
ratio for the decay into a pair of $W$ bosons (one of them being
virtual) is less than 10\%. The branching ratio for the interesting
decay into two photons will be at the level of BR($h^{\rm SM} \to \gamma
\, \gamma$) $\approx 2 \times 10^{-3}$.

The heavier non-singlet Higgs particles $h_3^0, a_2^0$ (and $h^\pm$)
will mainly decay into $b\bar b$ and $\tau^+ \tau^-$ ($tb$ and $\tau
\nu$)  pairs. The branching ratios will be 90\% and 10\% for 
the hadronic and leptonic decay modes, respectively. Due to the strong
enhancement of the $b$ and $\tau$ Higgs couplings as a result of the
large value of $\tan \beta \gsim 30$, the Higgs decays into SUSY
particles, as phase space allowed decays into pairs of the lighter
neutralinos and sleptons, will be strongly suppressed.  

The decays of the singlet-like neutral Higgs bosons (which are unlikely
to be produced as will be discussed later) are
induced through their (typically small) non-singlet components, implying
branching fractions corresponding to a SM-like Higgs.

Higgs-to-Higgs decays, which are a very peculiar possibility within the 
unconstrained NMSSM and which have been discussed in great detail in
Ref.~\cite{benchmark}, are generally absent in most of the parameter
space of the cNMSSM; only for the largest possible values of $\lambda$,
the lightest CP-even singlet-like Higgs mass can be sufficiently small
and its coupling to $h_2^0$ still sufficiently large as to give rise to
a non-negligible BR($h_2^0 \to h_1^0 \,h_1^0$). An illustrative example
is given by the following choice of input parameters
\begin{equation}
M_{1/2}=\,500\,{\rm GeV},\ 
m_0=\,46.5\,{\rm GeV},\ 
A_0=\,-142.65\,{\rm GeV},\ 
\tan \beta=\,31.72 ,\
\lambda =\, 0.0097\,,
\label{multi-b}
\end{equation}
where one obtains $m_{h_1^0}\simeq 50$ GeV and $m_{h_2^0}\simeq 118$ GeV.  
The dominant $h_2^0$ branching fractions are then 
\begin{equation}
{\rm BR}(h_2^0 \to b \,\bar b)\, =\, 64.1 \% \,, \quad 
{\rm BR}(h_2^0 \to h_1^0 \,h_1^0)\, =\, 12.1 \%\,.
\end{equation}
The state $h_1^0$ will mostly decay into $b\bar b$ (90\%) and $\tau^+
\tau^-$ (10\%) final states.

\subsection{Sparticle and Higgs production at the LHC}

Regarding sparticle production at the LHC, we summarize in
Table~\ref{tab:P1P3:sigma} the cross
sections for the most relevant production channels 
for the points P1 and P2 defined in Table~\ref{tab:b1b2}. 
These have been obtained by using 
the Fortran code Prospino \cite{prospino}, which calculates the cross
sections for sparticle production at hadron colliders to
next-to-leading order in perturbation theory, for the non-singlet sector
of the cNMSSM. 

The largest cross sections are expected for the strongly interacting 
production processes $pp \to \tilde q \tilde g$, $\tilde q \tilde q$
and $\tilde q \tilde q^*$, where $\tilde q$ denotes all squarks but the
stop. One obtains $\sigma_{\tilde q \tilde g} \simeq 0.67$~pb, 
$\sigma_{\tilde q \tilde q} \simeq 0.44$~pb and $\sigma_{\tilde q 
\tilde q^*} \simeq  0.22$~pb for the points P1 (and P1')  with
$M_{1/2}=500$~GeV, while for the point~P2 (likewise for P2')  with
$M_{1/2}=1000$ GeV, only $\sigma_{\tilde q \tilde q} \simeq 0.01$~pb, 
$\sigma_{\tilde q \tilde g} = 0.004$~pb and $\sigma_{\tilde q \tilde
q^*} = 0.002$~pb are expected. Assuming the LHC high luminosity option
of ${\cal L}= 100$~fb$^{-1}$, one obtains $10^4$ to
$10^5$ events for scenario P1, but only  $10^2$ to $10^3$ events  for
scenario P2 (before efficiency cuts are applied). The yield for gluino
pair production, $q\bar q/gg \to \tilde g \tilde g$,  is much smaller
as a result of the larger gluino mass.

As discussed in Section 4.1, all gluinos will decay into squarks.
Squarks decay into quarks and the wino-like neutralinos or charginos,
which then cascade mostly into the lighter $\tilde \tau$ NLSP. Hence
all sparticle  decay chains contain the $\tilde\tau_1$ NLSP, which will
finally decay into the singlino-like LSP, $\tilde \tau_1 \,
\to\,\chi_1^0 \,\tau$, possibly leading to a displaced
vertex. 

\begin{table}[!ht]\begin{center}
\renewcommand{\arraystretch}{1.2}
{\small 
\begin{tabular}{||l||c|c||}
\hline
\hline
$\sigma$ (pb) & \phantom{eV}P1\phantom{eV}& \phantom{eV}P2\phantom{eV}
\\\hline
$\tilde g\, \tilde g$ &  $9.5 \times 10^{-2}$ & $2.14 \times 10^{-4}$
\\\hline
$\tilde g\, \tilde q$ & $0.668$  & $4.28 \times 10^{-3}$
\\\hline
$\tilde q\, \tilde q$ & $0.436$  & $9.21 \times 10^{-3}$
\\\hline
$\tilde q\, \tilde q^*$ & $0.221$   & $1.64 \times 10^{-3}$
\\\hline
$\tilde t_1\, \tilde t_1^*$ & $3.69\times 10^{-2}$   & $2.63 \times 10^{-4}$
\\\hline\hline
$\tilde l_L\, \tilde l_L^*$ & $3.4\times 10^{-3}$  & $1.62\times 10^{-3}$
\\\hline
$\tilde l_R\, \tilde l_R^*$ & $1.17\times 10^{-2}$  & $8.87\times 10^{-4}$
\\\hline
$\tilde \nu_l\, \tilde \nu_l^*$ & $3.58\times 10^{-3}$  & $1.53\times 10^{-4}$
\\\hline
$\tilde \tau_1\, \tilde \tau_1^*$ & $4.8\times 10^{-2}$  & $3.46\times 10^{-3}$
\\\hline\hline
$\chi_2^0 \,\chi_2^0 $ &  $1.1 \times 10^{-3}$ &$6.22 \times 10^{-5}$
\\\hline
$\chi_2^0 \,\chi_3^0 $ &  $1.73 \times 10^{-4}$ &$8.67 \times 10^{-6}$
\\\hline
$\chi_2^0 \,\chi_1^\pm $ &  $5.37 \times 10^{-4}$ &$6.53 \times 10^{-5}$
\\\hline
$\chi_3^0 \,\chi_3^0 $ &  $1.79 \times 10^{-3}$    &$5.74 \times 10^{-5}$
\\\hline
$\chi_3^0 \,\chi_1^\pm $ &  $6.51 \times 10^{-2}$    &$7.49 \times 10^{-3}$
\\\hline
$\chi_1^+ \,\chi_1^- $ &  $3.53 \times 10^{-2}$     &$1.17 \times 10^{-3}$
\\\hline \hline
\end{tabular}}\end{center}\vspace*{3mm}
\vspace*{-3mm}
\caption{Production cross sections (in pb) for strongly (upper part) 
and weakly (lower part) interacting sparticles at the LHC 
in the points P1 and P2 defined in Table~\ref{tab:b1b2}, as obtained
with the program  Prospino~\cite{prospino}. Here $\tilde q$ denotes all
squarks but the  stop, and $l = \mu$ or $e$.
}\label{tab:P1P3:sigma} 
\end{table}

The simplest possible squark decay cascades at the LHC -- originating
mostly from right-handed squarks of the first generation $\tilde q_R$ --
are thus of the form
\begin{equation}\label{qr:decays}
\tilde q_R \ \to \  q\, \chi_2^0 \,   \
\begin{array}{lcll}
 \nearrow & q \tilde \tau_1 \tau  &  \rightarrow \ q \tau \tau
  \chi_1^0  & \quad (\mathrm{P1}:88\%;\;\mathrm{P2}:74\%)\\
  \searrow & q \tilde l_R l \to q l \tilde \tau_1 \tau  
  &  \rightarrow \ q  \tau \tau \chi_1^0+\ell  
  & \quad  (\mathrm{P1}:12\%;\;\mathrm{P}2:25\%) 
\end{array}
\end{equation}
where, according to Table~\ref{tab:P1P3:decay},   
the first case occurs $\sim 88\%$ for point P1 and $\sim 74\%$ for
point P2. The second case occurs only $\sim 12\%$ and  $\sim 25\%$ 
for P1 and P2, respectively. These cascades 
typically lead to events with 3 jets per $\tilde q_R$ (one hard quark
and two tau jets, also potentially hard, depending on the momentum of the
decaying $\chi_2^0$ neutralino), but the leptonic decays of the  $\tau$
can possibly also be used.

As can be seen from eq.~(\ref{qr:decays}), right-handed squark 
cascade decays containing a lepton in the final state are clearly 
sub-dominant. 
In the case of left-handed squarks, the decays would lead to more $e,
\mu$ lepton final states, since $\tilde q_L$ mainly decays into $q
\chi_3^0$ and $q \chi_1^\pm$. The wino-like neutralino $\chi_3^0$ and
chargino $\chi_1^\pm$ decay either directly to $\tilde \tau_1$ and
$\tilde  \nu_\tau$ (the latter decaying predominantly into $\tilde
\tau_1 W$) or to $\tilde l_L,  \tilde \nu_l$. These then decay
into $l \chi_2^0$ and $\nu \chi_2^0$, respectively. The bino-like
$\chi_2^0$ would  dominantly lead to $\tilde \tau_1 \tau$ and, to a minor
extent, to $\tilde l_R l \to l l \tilde \tau_1 \tau$ final states.
Therefore, one has a  non-negligible probability for one,
two, three and even four $e,\mu$ leptons in the final states. For
instance, the branching ratios into the four lepton topology would be
$\approx 1\%$ and $\approx 2\%$ for, respectively, points P1 and P2.   

For chargino and neutralino or mixed pair production, the cross sections  
are also shown in the lower part of Table \ref{tab:P1P3:sigma}; they  are
sizable enough only for point P1, where the phase space is not too 
penalizing. In the case of $\chi_1^\pm \chi_1^\mp$ and $\chi_3^0
\chi_1^\pm$, they lead to a few thousand events for an integrated 
luminosity  of
${\cal L}= 100$ fb$^{-1}$, before cuts are applied.  However,
since charginos and neutralinos also dominantly cascade into $\tilde
\tau$ lepton final states, only a very small fraction leads to the nice
signature of multi $e, \mu$ lepton events. 

The cross sections for the
Drell-Yan production of slepton pairs is of the same order as the one of
charginos and neutralinos, the highest one being $p \bar p \to \tilde
\tau_1 \tilde \tau_1^*$ pair production (with $\sigma \approx 50$ fb for
P1, where $m_{\tilde \tau_1} \approx 130$ GeV) as a result of the more
favorable phase space\footnote{In fact, at the Tevatron, 
the only process which might have a
significant production cross section is the pair production of the lighter 
stau: one would have  $\sigma( p\bar p \to  \tilde \tau_1 \tilde \tau_1^*) 
\approx 15$ fb for  $m_{\tilde \tau_1} \approx 100$ GeV. The final state 
would then essentially consist of missing energy and two tau final states, 
the latter potentially decaying into muons with large impact parameters, 
since the $\tilde \tau_1$ lifetime can be e.g. of the order of 20 ps for  
$m_{\tilde \tau_1} \approx 100$ GeV and $\lambda \approx 10^{-5}$. Such 
events would share some (but definitely not all) of the peculiarities of 
the displaced multi-muon events recently reported by (part of) the CDF 
collaboration \cite{CDF-muons}. However, in our case only a handful of 
events would have been produced 
and the $\tau$ multiplicity would be far too small.}. 
 
Since cascade decays leading to $e,\mu$ lepton final states will be
generally rare in the cNMSSM, the measurements of the sparticle masses
from kinematical endpoints and, thus, the determination of  some of the
soft--SUSY breaking parameters, cannot be performed with the same level of
accuracy as it would be the case for the MSSM. Note also that, in most
cases, one of the leptons originates from the three-body $\tilde l_R \to 
l \tilde \tau_1 \tau$ decay. 

Hence, a SUSY signal can certainly be observed at the LHC by
looking, for instance, for final states with hard jets plus a large
amount of missing energy.
However, precision measurements through the endpoints
of decay spectra will be more complicated to perform in the case of
$\tau$ final states which are overwhelming in the cNMSSM. 

Finally, let us make a few comments on Higgs detection in
this scenario. As previously discussed, in most cases the cNMSSM Higgs
sector reduces to the one of the MSSM in the decoupling regime. 
The SM-like CP-even Higgs boson with its mass in the
115--120~GeV range can be discovered first at the Tevatron (if
enough integrated luminosity is collected) in the Higgs--strahlung
process  $pp \to W+$Higgs leading to $ l\nu b\bar b$ final states.
At the LHC, the most relevant production and decay processes will be the
gluon-gluon and vector boson fusion, $gg \to$ Higgs and $qq \to 
qq+$Higgs, with the Higgs boson decaying into two photons (and possibly
$\tau^+\tau^-$ final states in the vector boson fusion process).

We recall the phenomenon of the ``cross-over'' for certain values of 
$M_{1/2}$ and $m_0$, 
which
corresponds to two nearly degenerate CP-even Higgs bosons sharing their
couplings to the SM gauge bosons, quarks and leptons. This is the case
illustrated by the last point P2' of Table~\ref{tab:b1b2},
where the lightest Higgs bosons $h_1^0$ and $h_2^0$ have very similar
singlet/doublet components $\sim 0.7$, i.e. similar production cross
sections and branching fractions, and a mass splitting of about
$10$~GeV. In principle, both states could be
observed\footnote{The situation here is similar to the MSSM intense
coupling regime discussed in Ref.~\cite{intense} in which all neutral
MSSM bosons are close in mass; however, in the cNMSSM, the couplings of
the lighter CP-even states  to down-type fermions are not particularly
enhanced.}.
For smaller values of $\lambda$, the mass splitting in the
cross-over region can be considerably smaller, less than 1~GeV for
$\lambda \lsim 10^{-4}$.  Then, the sum of both Higgs bosons gives rise
to a single SM-like Higgs boson peak, which can be quite difficult to
resolve at the LHC.

The heavier non-singlet Higgs particles have significant couplings
to SM down-type fermions in view of the large values of $\tan\beta 
\gsim 30$. They can be observed in associated production with $b\bar b$ pairs
and decays into $\tau^+ \tau^-$ pairs for the neutral particles, and
associated production with $t b$ pairs with decays into $\tau \nu$ in the
case of charged Higgs bosons. However, this  is only possible for low
values of $M_{1/2} \lsim 700$ GeV, which lead to light enough Higgs
particles with sufficient production cross sections. In any case, a very
high luminosity is required to observe these particles. 

Apart from the ``cross-over'' region, the singlet-like Higgs states are
generally inaccessible as they couple too weakly to SM fermions and
gauge bosons. Only if the mass of a singlet-like CP-even Higgs boson
$h_1^0$ is sufficiently small and $\lambda$ sufficiently large, as is 
the case for the
point given in eq.~(\ref{multi-b}), the branching ratio $h_2^0 \to h_1^0
h_1^0$ of the then SM-like $h_2^0$ can be sizable (leading to a
slightly reduced $h_2^0 \to \gamma \gamma$ branching ratio). Whether
this production of $h^0_1$ leading to difficult $4b, 2\tau 2b$ and
$4\tau$ final state topologies could be detected \cite{benchmark}, needs
also a dedicated detailed investigation.

\subsection{Searches at the ILC}

Due to the relatively heavy Higgs and sparticle spectrum, a
multi--TeV $e^+ e^-$ collider would be required to produce all the
states of the constrained NMSSM. At a 500 GeV ILC, only the two lighter
neutralinos and the right-handed sleptons can be produced even for $M_{1/2}
\lsim 500$ GeV. The production of wino-like charginos and neutralinos as
well as the left-handed  sleptons  would need a larger center of mass
(c.m.) energy of $\approx 1$ TeV.  The detection of the various SUSY states
would be straightforward in the clean environment of the ILC. In
particular, a $\tilde \tau_1$ nearly degenerate with the LSP
can be detected, as shown in detailed simulations for somewhat similar
MSSM scenarios \cite{stau-ILC,DCR-ILC}.  The masses of the SUSY
particles could be accurately determined, at least through threshold
scans, as has been discussed in detail in  Ref.~\cite{DCR-ILC}, and a
clear distinction between the cNMSSM and other scenarios could be made. 

As far as the Higgs sector is concerned, only the lighter CP--even and
CP--odd Higgs states would be kinematically accessible at an ILC with 
a c.m. energy less than 1 TeV. The SM-like Higgs particle can be easily
detected and its properties probed in detail, the Higgs mass range 
around  $\approx 120$ GeV being the ideal one for a 500 GeV ILC
\cite{DCR-ILC}. The scenario with sizable singlet/doublet mixing
between the $h_1^0$ and $h_2^0$ states can be probed in the
Higgs-strahlung process, $e^+ e^- \to Z+$Higgs, where the separate Higgs
states can be  disentangled even if they are nearly degenerate in mass,
as the resolution on the Higgs masses in this process is smaller than
100 MeV \cite{DCR-ILC}. The scenario of eq.~(\ref{multi-b}), in which
there is a light CP-even Higgs particle, allows the $h_2^0 \to
h_1^0 h_1^0$ decay  to occur, which can also be probed in the Higgs-strahlung
process $e^+ e^- \to h_2^0 Z \to \mu^+ \mu^- b\bar b b\bar b$,
and both  $h_1^0$ and $h_2^0$ masses could be accurately determined. The
singlet-like  state $a_1^0$ could be also accessible in the pair
production process $e^+ e^- \to h_2^0 a_1^0 \to b\bar b b\bar b$ unless 
the $a_1^0$ mass is too large or the coupling $\lambda$ prohibitively
tiny.

\subsection{Direct and indirect detection of dark matter}

For completeness, let us comment on the prospects for detecting the
singlino-like LSP $\chi^0_1$ (with non-singlet components of ${\cal O}
(\lambda$)) in astroparticle experiments. For the direct LSP detection 
via $\chi^0_1$-nuclei interactions, the prospects  are quite dim: due to
the very small values of $\lambda$ that are required from compatibility
of the Higgs sector
with LEP bounds, one finds extremely small WIMP-nucleon
cross sections\footnote{In different
studies~\cite{old:dm,Balazs:2008ph}, large direct detection cross
sections have been obtained for the case of the non-constrained or
semi-constrained NMSSM. The large cross sections were associated to
lighter $\chi^0_1$ and larger values of $\lambda$, which are not
possible in the  present case.}. We computed these cross sections using
the 
recent version of the dark matter code
MicrOMEGAS~\cite{Belanger:2008sj}, adapted to the
NMSSM. The result is that the relevant values for the spin-independent  and
spin-dependent  proton and neutron interaction cross-sections are always
below $\sim 10^{-13}$~pb, implying an expected number of events smaller
than $\mathcal{O}(10^{-8\div -9})$/day/kg for both $^{73}$Ge and
$^{131}$Xe nuclei, and less than  $\mathcal{O}(10^{-11\div -12})$/day/kg
for $^{3}$He. 

Furthermore, since $\chi_1^0$ has extremely small couplings  to
SM fermions and gauge bosons, the annihilation cross sections of
$\chi_1^0$ into SM particles are always extremely small. The only 
significant channel could be the annihilation process $\chi_1^0 \chi_1^0
\to a_1^0$ with the CP-odd singlet-like $a_1^0$ state being sufficiently
light to be produced on--shell. However, the $a_1^0 \chi_1^0 \chi_1^0$
coupling is proportional to $\kappa$ which is tiny in our case,
$\kappa \lsim 10^{-3}$. 

Thus all cross sections involving the singlino LSP are extremely small; 
in other words, the fully constrained cNMSSM can be excluded by the
direct or indirect detection of a WIMP-like dark matter candidate.

\section{Summary and conclusions}

Due to its simplicity, namely a scale invariant superpotential and
universal SUSY-breaking terms at the GUT scale, the cNMSSM is a very
attractive supersymmetric extension of the standard model. We have
found that for small values of the NMSSM specific 
Yukawa coupling, $\lambda
\lsim 10^{-2}$, the cNMSSM can satisfy all present phenomenological
constraints including LEP constraints on the Higgs sector. The correct
dark matter relic density is obtained for small $m_0$ and $A_0$
as compared to $M_{1/2}$. Moreover, in the region $M_{1/2} \approx
550 - 600$~GeV, the deviation of $(g-2)_\mu$ from its SM model value
and the two excesses of Higgs-like events observed at LEP can be
simultaneously explained.

The cNMSSM sparticle spectrum allows to discriminate it from most versions
of the MSSM: all squarks are lighter than the gluino, and due to the
weakly coupled singlino-like LSP and the stau NLSP, the latter will show
up in practically all sparticle decay cascades.  
In some regions of the parameter space, the stau lifetime can be sufficiently 
large, possibly leading to visibly displaced vertices.

The mass of the SM-like CP-even Higgs boson is constrained to lie in the
115 -- 120~GeV range, and for certain regions in parameter space it could
strongly mix with the singlet-like Higgs state. In a small -- different --
region of the parameter space, decays of the SM-like CP-even Higgs boson
into two singlet-like Higgs states are possibly detectable at an ILC.

Note that the model is very predictive: as it becomes clear from 
Figs.~\ref{fig:m12:phenoX0st} and \ref{fig:m12:phenoGQL}, the
measurement of one sparticle mass (or mass difference) would allow to
predict quite accurately the complete remaining sparticle spectrum. On
the other hand, due to the singlino-like LSP together with small
$\lambda$, dark matter detection signals can rule out the present
model.

Hopefully, the cNMSSM can be tested in the near future at the LHC. To
this end the sensitivities of the ATLAS and CMS detectors to sparticle
decay cascades involving the stau NLSP should be thoroughly studied.
\bigskip

\noindent {\bf Acknowledgements} \smallskip

\noindent We acknowledge  support from the French ANR project PHYS@COL\&COS and
discussions with S.F. King, S. Moretti and W. Porod. AD is grateful to  the
Alexander von-Humboldt Foundation (Germany) and to the theory group in
Bonn for the hospitality extended to him.

\end{document}